\documentclass[showpacs,prd,twocolumn,eqsecnum,superscriptaddress,a4paper,preprintnumbers]{revtex4}

\pdfoutput=1

\usepackage{amsfonts}
\usepackage{amsmath}
\usepackage{bm}

\usepackage{graphicx}
\usepackage{subfigure}
\graphicspath{{EPS/}}

\usepackage{ifpdf}
\ifpdf\usepackage[pdftitle={GW radiometry: Mapping an SGWB},%
colorlinks,citecolor=red,linkcolor=blue]{hyperref}\fi
\ifpdf\usepackage{epstopdf}\else\fi

\def \Om{\mathbf{\hat{\Omega}}}

\def \tp{{t^\prime}}

\def \Dt{{\Delta t}}

\def \d{\mathrm{d}}

\def \ft{\widetilde}
\def \Dx {\mathbf{\Delta x}}
\def \Ddx {\mathbf{\Delta \dot{x}}}
\def \x{\mathbf{x}}
\def \Omp{\mathbf{\hat{\Omega}}'}
\def \DOm{\mathbf{\Delta \Omega}}
\def \dom {\Delta \Omega}
\def \P {\mathcal{P}}
\def \ncone {\mathbf{\hat{n}}_{\text{cone}}}

\def \snc{\mathrm{sinc}}

\def \Om {{\bf \hat{\Omega}}}

\def \R{{\cal R}}
\def \h {{1 \over 2}}
\def \nrm {\parallel}
\def\be{\begin{equation}}
\def\ee{\end{equation}}
\def\bea{\begin{eqnarray}}
\def\eea{\end{eqnarray}}
\def \no {\nonumber}
\def\lsim{\mathrel{\rlap{\lower4pt\hbox{\hskip1pt$\sim$}}
    \raise1pt\hbox{$<$}}}                
\def\gsim{\mathrel{\rlap{\lower4pt\hbox{\hskip1pt$\sim$}}
    \raise1pt\hbox{$>$}}}                

\def \Om {\hat{\mathbf{\Omega}}}
\def \T {\Delta t}
\def \d {\mathrm{d}}
\def \ft {\widetilde}
\def \Dx {\mathbf{\Delta x}}
\def \DOm {\mathbf{\Delta\Omega}}
\def \P {\mathcal{P}}
\def \ncone {\hat{\mathbf{n}}_\text{cone}}
\def \Npix {N_\text{pix}}
\def \Pest {\hat{\bm{\P}}}
\def \i {\mathfrak{i}}
\def \Nb{N_{\mathrm{b}}}

\begin{document}

\preprint{IUCAA 28/07}
\preprint{LIGO-P070033-Z}

\title{Gravitational wave radiometry: Mapping a stochastic gravitational wave background}

\author{Sanjit~Mitra}\email{sanjit@oca.eu}
\affiliation{Inter-University Centre for Astronomy and Astrophysics,\\
Post Bag 4, Ganeshkhind, Pune 411007, India}
\affiliation{Observatoire de la C\^ote d'Azur, BP 4229, 06304 Nice Cedex 4, France}

\author{Sanjeev~Dhurandhar}\email{sanjeev@iucaa.ernet.in}
\affiliation{Inter-University Centre for Astronomy and Astrophysics,\\
Post Bag 4, Ganeshkhind, Pune 411007, India}

\author{Tarun~Souradeep}\email{tarun@iucaa.ernet.in}
\affiliation{Inter-University Centre for Astronomy and Astrophysics,\\
Post Bag 4, Ganeshkhind, Pune 411007, India}

\author{Albert~Lazzarini}\email{lazz@ligo.caltech.edu}
\affiliation{LIGO Laboratory, California Institute of Technology, \\
MS 18-34, Pasadena, CA 91125, USA}

\author{Vuk~Mandic}\email{vmandic@ligo.caltech.edu}
\affiliation{LIGO Laboratory, California Institute of Technology, \\
MS 18-34, Pasadena, CA 91125, USA}

\author{Sukanta~Bose}\email{sukanta@mail.wsu.edu}
\affiliation{Department of Physics, Washington State University, Pullman, WA 99164-2814, USA}

\author{Stefan~Ballmer}\email{sballmer@caltech.edu}
\affiliation{LIGO Laboratory, California Institute of Technology, \\
MS 18-34, Pasadena, CA 91125, USA}

\begin{abstract}

The problem of the detection and mapping of a stochastic gravitational
wave background (SGWB), either cosmological or astrophysical, bears a
strong semblance to the analysis of the cosmic microwave background
(CMB) anisotropy and polarization, which too is a stochastic field,
statistically described in terms of its correlation properties. An
astrophysical gravitational wave background (AGWB) will likely arise from an
incoherent superposition of unmodelled and/or unresolved sources and
cosmological gravitational wave backgrounds (CGWB) are also predicted in certain
scenarios.  The basic statistic we use is the cross-correlation
between the data from a pair of detectors. In order to `point' the
pair of detectors at different locations one must suitably delay the
signal by the amount it takes for the gravitational waves (GW) to travel to both detectors
corresponding to a source direction. Then the raw (observed) sky map of the SGWB
is the signal convolved with a beam response function that
varies with location in the sky.  We first present a thorough analytic
understanding of the structure of the beam response function using an
analytic approach employing the stationary phase approximation. The
true sky map is obtained by numerically deconvolving the beam function
in the integral (convolution) equation.  We adopt the maximum likelihood
framework to estimate the true sky map using the conjugate gradient method
that has been successfully used in the broadly similar, well-studied
CMB map making problem.  We numerically implement and demonstrate the
method on signal generated by simulated (unpolarized) SGWB
for the GW radiometer consisting of the LIGO pair of detectors at Hanford and Livingston.
We include `realistic' additive Gaussian noise in each data
stream based on the LIGO-I noise power spectral density.  The
extension of the method to multiple baselines and polarized GWB is
outlined.  In the near future the network of GW detectors, including the
Advanced LIGO and Virgo detectors that will be
sensitive to sources within a thousand times larger spatial volume,
could provide promising data sets for GW radiometry.

\end{abstract}

\maketitle
\section{Introduction}
\label{intro}

The existence of gravitational waves (GW), has long been verified `indirectly'
through the observations of Hulse and Taylor~\cite{HT}. However,
direct observation of such waves with manmade gravitational wave detectors has
been lacking.  At present the laser interferometric
detectors have achieved sensitivities close to that required for
detecting such waves~\cite{GBGWD}. The space mission LISA~\cite{lisa}
is also planned by the ESA and NASA to detect low frequency GW.  The
significance of the direct detection of GW lies, not only in the
opening of an entirely new window into observational astronomy by
probing phenomena in the regime of strong gravity; it further
promises to test our present theories of gravitation.

Different types of GW sources have been predicted and may be directly
observed by Earth-based detectors in the near future
(see~\cite{Thorne87,Thorne95,Fl_GR15,Schutz99,Thorne,Kalogera,schutz}
and references therein for recent reviews): (i)~Transient sources -- such
as binary systems of neutron stars (NS) and/or black holes (BH) in
their in-spiral phase, BH/BH and/or BH/NS mergers, and supernovae
explosions, whose signals last for a time much shorter, typically
between a few milli-seconds and a few minutes, than the planned
observational time; (ii)~Continuous wave (CW)
sources -- e.g. rapidly rotating neutron stars, where a weak
deterministic signal is continuously emitted, and (iii)~Stochastic backgrounds of radiation, either of primordial or astrophysical origin.
 
In this paper we will address the problem of a spatially resolved search of 
the gravitational wave stochastic background. This approach was advocated in the LIGO
technical note~\cite{LazzWeiss} and the basic analysis was recently
implemented on the fourth science run data from the LIGO interferometers
to prepare an upper limit map~\cite{S4}.
Our main focus will be on a stochastic astrophysical GW background (AGWB), 
which might arise from a superposition of a large number of independent and unresolved GW
sources.  The gravitational wave background can arise from a variety
of sources: supernovae with asymmetric core collapse, 
binary black hole (BBH) mergers, GWs from low-mass X-ray binaries (LMXBs) and hydrodynamical instabilities in neutron stars (r-modes),
or even GWs from astrophysical objects that 
we never knew existed. When a collection of any subset of these
sources is unresolvable, it can appear as a stochastic GW background
(SGWB) of a variable duration in our detectors of interest.  
While an astrophysical background will provide information about
our immediate neighborhood, cosmological GW backgrounds (CGWB)
could probe the physics of the early universe.  There exist
cosmological scenarios (e.g., cosmic strings and super-string models)
which predict CGWB that should be
detectable by Advanced LIGO~\cite{Maggiore:1999vm}.

We propose and develop a data analysis method that measures and
maps the power in the SGWB from a specific location in the sky - GW
radiometry using a network of detectors. We find that the angular
resolution essentially depends on the effective GW bandwidth and the
linear size of the network. In this paper we will restrict ourselves
to the network of the two 4km LIGO detectors at Hanford (LHO) and Livingston (LLO). 
For the purposes of our analysis, we take their noise curves to be identical 
with the LIGO-I design power spectral density~\cite{LigoInoise}. Our future plan is 
to include VIRGO and other detectors around the world in the
numerical implementation of this analysis.

The basic statistic is the cross-correlation between the data
from a pair of detectors. In order to `point' the pair of detectors at
different locations one must suitably delay the signal by the amount
it takes for the GW to travel to both detectors corresponding to the
source direction. This delay will be a function of the source position
and will vary as the Earth rotates. Using the delay allows the
detectors to sample the same wavefront from the source. The cross-spectrum
formulation has been carried out in~\cite{crsfrm,flan}. Methods for
searching for isotropic backgrounds~\cite{alnrmn} using the
cross-correlation and for anisotropies using spherical harmonic
decomposition~\cite{ottaln} have been devised. Efforts have also been made
to devise methods to measure the spherical harmonic moments of the SGWB anisotropy 
using a network of ground or space based detectors~\cite{cornish, Taruya}.
Here we focus on a spatially resolved search and the final goal 
is to make a map of the true SGWB sky. We achieve this goal by pixelizing the sky, 
that is, we use a pixel basis.

The advantage of a spatially resolved search is seen immediately if we examine the so called overlap reduction factor, which partially determines the fractional power of source spectrum the search filter will receive at different frequency bands. The overlap reduction factor, normally denoted by $\gamma (f)$ in the literature~\cite{alnrmn} for the isotropic unpolarized background, becomes a time-dependent factor $\gamma(\Om, f, t)$ for the spatially resolved search. For the LIGO detectors, $\gamma (f)$ quickly reduces to zero beyond few tens of Hz, while $\gamma(\Om, f, t)$ has infinite bandwidth. So the bandwidth of the spatially resolved search is essentially detector bandwidth limited. This is typically valid for a network of detectors and therefore important from the point of view of the sensitivity regime of GW detectors which lies in this region.

As in radio-interferometry, the correlation statistic so
constructed produces a `dirty' map where a point source does not
produce a point image, but one that is smeared by a beam response
function (beam, for brevity).  The `cleaned' GW sky map is obtained
from the measured cross-correlation statistic by deconvolving the
beam. In other words, to obtain the GW power from each direction in
the sky one needs to solve an integral equation where the measured
power (data) is a convolution of the actual power with a kernel
(beam). In order to understand the structure of the beam we carry out a numerical and
an analytical study using the stationary phase approximation (SPA). We find
that at low declinations (latitudes) of a point source, the kernel essentially has
the shape of a figure `8' with a bright spot at the intersection. The
bright spot is at the location of the point source. The figure of `8'
continuously changes and bifurcates into a `tear drop' as the point
source moves to higher declinations. The declination at which this
bifurcation occurs is determined by the half-angle of the cone traced
out by the vector joining the two detectors. For the LIGO detectors
this declination is about $26^{\circ}$. The size of the bright spot or
effective sky patch, defined by a certain percentage of reduction in
the beam response function, say $50\%$, is determined by the inverse of the band-width
divided by the light (GW) travel time between the
detectors. Considering a broadband source and LIGO detectors having
kHz bandwidth with 10 ms light travel distance between them, the
angular size is about $5^\circ$ in radius. We find that these results
agree very well with those obtained by applying singular value
decomposition to the kernel matrix; the eigenvalues fall off steeply
after a certain point which determines essentially the number of
`degrees of freedom' of the kernel matrix and thereby the size of the
sky patch.

We employ the maximum likelihood (ML) approach for deconvolving the
sky map. The integral equation for a discrete pixelised sky leads to a
set of linear algebraic equations. Several deconvolution algorithms
exist in literature for solving such a problem. However, because of
the broad similarity of our present problem with the cosmic microwave
background (CMB) analysis, we have opted for techniques that have been
successfully applied to deconvolve CMB sky maps. Moreover, the ML approach
provides a framework to study the SGWB anisotropy in other basis of interest,
e.g., the spherical harmonic basis. We find the ML
estimate by employing the conjugate gradient method.  To verify our
method, we apply the analysis on simulated sky maps mapped by a GW radiometer
consisting of LIGO pair of detectors LHO and LLO. We generate `realistic'
colored Gaussian noise corresponding to LIGO-I design sensitivity
curve~\cite{LigoInoise} detector noise and embed various simulated sky maps of the GW stochastic background into the noise. We demonstrate that the true sky maps can be recovered satisfactorily.

The paper is organized as follows: In section~\ref{GWRd} we briefly
review the GW radiometer concepts and obtain an expression for the GW
radiometric cross-correlation signal, which is then optimized for the
maximum signal-to-noise (SNR).  In section~\ref{sec:convEq}, we set up the integral
equation that must be solved in order to obtain the true sky map from
the data. A time-frequency analysis is performed and the directed optimal filter is derived for anisotropic searches. Moreover, a stationary phase analysis is presented which provides us with the understanding of the kernel or the point spread function. In section~\ref{sec:ML}, we describe the maximum likelihood approach and the conjugate gradient method and apply it to simulated data to test the efficacy of this method. Later subsections outline the extension of
the GW radiometer analysis to incorporate multiple baselines obtained with a network of
detectors and the extension of the GW radiometer to search for polarized SGWB. The numerical implementation of the method is described in section~\ref{sec:numRes}. We conclude in section~\ref{remarks}.

\section{GW radiometer employing  Earth rotation aperture synthesis}
\label{GWRd}

\subsection{The principle of a radiometer}
\label{princ}

\begin{figure}
\centering
\includegraphics[width=0.45\textwidth]{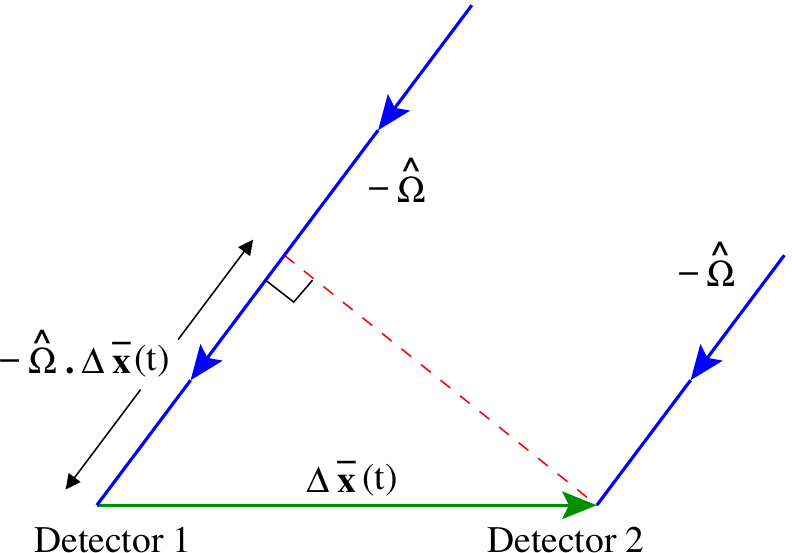} 
\caption{Geometry of an elementary radiometer. Above, ${\Dx} (t)$ is the separation or baseline vector between the two detectors; as the Earth rotates, its direction changes, but its magnitude remains fixed. The direction to the source $\Om$ is also fixed in the barycentric frame. The phase difference between signals arriving at two detector sites from the same direction is also shown.}
\label{radiometer}
\end{figure}

Radiometry or aperture synthesis is a well known technique in
radio astronomy and CMB experiments. The idea is to \emph{point} a pair of
detectors separated by a baseline to a desired direction in the sky
by  introducing an appropriate 
time-delay between their data-streams. This delay corresponds to the
difference between the times of arrival of a GW signal if it were to 
arrive at those two detector sites from that direction. For
a given source in the sky, this delay will change as the baseline orientation
changes due to the rotation of the earth.
The cross-correlation of the data from the two detectors, 
appropriately time-delayed, would cause potential GW signals arriving 
from the chosen direction to interfere constructively. Whereas signals from
other directions will tend to cancel out because of destructive
interference. This principle of Earth rotation aperture synthesis,
which is well-known in radio astronomy, could very well be used in GW 
astronomy using pairs of GW antennae.
Figure \ref{radiometer} illustrates the principle on which
the GW radiometer works.
   
We consider the Celestial Equatorial frame whose origin coincides with the
centre of the Earth. The axes are defined as follows: For a fixed but
arbitrarily chosen time $t=0$, the $x$-axis is directed towards the 
intersection of the equator and the longitude $\phi = 0$, which can be
taken as the Greenwich meridian; the $z$-axis is directed towards the 
North Celestial Pole and the $y$-axis is chosen
orthogonal to the previous two axes forming a right-handed triad. The
Earth rotates in this frame with the angular velocity $\omega_E = 2 \pi/(1~{\rm sidereal~day}) \sim 7.3 \times 10^{-5}$ radians/sec oriented along
the $z$-axis. The two detectors are at locations $\x_I$ (where $I = 1,2$), 
and the baseline vector joining the two sites is $\Dx := \x_2 - \x_1$. $\Om$ is the unit 
vector in the direction to the source, which is fixed in the barycentric frame.
The baseline $\Dx$ rotates with the Earth, but its
magnitude $|\Dx|$, which is the distance between the
detectors, remains constant. The map of the SGWB can be constructed
by performing this synthesis for each location in the sky, patch by
patch. An approximate size of the patch or resolution of the GW radiometer and
number of patches required to cover the full sky can be estimated from
the following simple argument. This simple model produces a fringe
pattern with resolution given by the standard 
formula for the central width $\Delta \theta \sim \lambda_\text{GW} /
(|\Dx| \sin \theta)$ where $\lambda_\text{GW}$ is the GW wavelength
and $\theta$ is the angle of incidence. This is, however, a naive estimate.
A better estimate of the resolution would follow from considerations involving
the pixel-to-pixel Fisher information matrix. In which case
the solid angle resolution will scale inversely proportional to the square of the SNR.

In this paper we will consider a GW radiometer made using the two $4$~km LIGO detectors 
LHO and LLO in the United States. However, this
analysis is equally well applicable to any other baseline or a network of baselines
involving other
detectors such as VIRGO, TAMA, GEO etc. The light travel time between
the LHO and LLO is $\sim 10$ ms, which at a bandwidth of 1 kHz 
gives a resolution $\Delta \theta$ of few degrees. This resolution
implies that a few
thousand patches or pixels are required to completely cover the full
sky. This estimate, in turn, will be useful for assessing the computational 
cost and numerical complications in handling large matrices
for obtaining the full skymap. We make these estimates more robust below.

\subsection{Basic Framework and the Statistical Properties of the Data}
\label{signal}

We first set up the notation and framework needed for investigating
the problem.  

In the transverse traceless (TT) gauge the metric perturbation
$h_{ij}(t,\mathbf{x})$ in the equatorial frame can be expanded in
terms of plane waves of the two polarizations $+,\times$ in the
following way:
\begin{equation}
h_{ij}(t,\mathbf{x}) = \int_{-\infty}^\infty \d f \int_{S^2} \d\Om \, \ft{h}_A(f,\Om) \, e^A_{ij}(\Om) \, e^{2\pi i f (t + \Om\cdot\mathbf{x}/c)}, 
\label{eq:plwave}
\end{equation}
where a tilde on a variable denotes its Fourier transform; the complex Fourier amplitudes satisfy the relation $\ft{h}_A^*(f,\Om) \ = \ \ft{h}_A(-f,\Om)$ owing to the reality of $h_{ij}(t,\mathbf{x})$. 
Here $A = \{+, \times \}$ is the polarization index  and summation over the repeated index $A$ is implied. In terms of the spherical polar coordinates $(\theta,\phi)$, the source
direction is given by
\begin{equation}
\Om = \cos\phi \, \sin\theta \, \hat{\mathbf{x}} \ + \ \sin\phi \, \sin\theta \, \hat{\mathbf{y}} \ + \ \cos\theta \, \hat{\mathbf{z}} \,
\end{equation}
and, hence, the infinitesimal solid angle along the direction $\Om$ is
$\d\Om \equiv \sin\theta\,\d\theta\,\d\phi$.
Note that the wave propagation direction is $-\Om$.
The polarization tensors $e^A (\Om)$ are defined by the following expressions:
\begin{subequations}
\begin{eqnarray}
e^+(\Om) &=& \hat{\mathbf{e}}_{\theta} \otimes \hat{\mathbf{e}}_{\theta} - \hat{\mathbf{e}}_{\phi}\otimes \hat{\mathbf{e}}_{\phi} \,, \\
e^\times (\Om) &=& \hat{\mathbf{e}}_{\theta} \otimes \hat{\mathbf{e}}_{\phi} + \hat{\mathbf{e}}_{\phi} \otimes \hat{\mathbf{e}}_{\theta} \,,
\end{eqnarray}
\end{subequations}
where 
\begin{subequations}
\begin{eqnarray}
\hat{\mathbf{e}}_{\theta} &=& \cos\phi \cos\theta \, \hat{\mathbf{x}}  +  \sin\phi \, \cos\theta \, \hat{\mathbf{y}} \ - \ \sin\theta \, \hat{\mathbf{z}} \,, \nonumber \\
\hat{\mathbf{e}}_{\phi} &=& - \sin\phi \, \hat{\mathbf{x}} \ + \ \cos\phi\, \hat{\mathbf{y}} \,,\nonumber
\end{eqnarray}
\end{subequations}
are the two orthonormal basis vectors on a two-sphere.

The statistics of the GW signal can be best described in the Fourier domain: 
If we assume the signal to be stochastic and uncorrelated in the two 
polarizations\footnote{It is
important to note that, even if the polarizations are independent in
certain direction, there can be mixing between polarizations in other
directions due to the rotation of basis vectors. This complication has
not been considered here, although it may not be too difficult to
incorporate this in our analysis.}, different frequencies and different directions, then
the Fourier components of the GW strain obey,
\bea
&& \langle \ft{h}_A^*(f,\Om) \, \ft{h}_{A'}(f',\Om')
\rangle \nonumber \\ && \quad = \ \delta_{AA'} \, \delta(f-f') \,
\delta^2(\Om-\Om') \, \P_A(\Om) \, H(f) \,,~~~~~~
\label{eq:indGWBmodes}
\eea 
where $\P_A(\Om)$ is proportional to the strength of the SGWB in
the direction $\Om$ and $H(f)$ is the {\em two sided} GW source Power
Spectral Density (PSD).
The interpretation of the quantity
$\P_A(\Om)$ can be made apparent by relating it to the {\em specific intensity}
or {\em brightness}~\cite{RL} of GW $I_\text{GW}(f,\Om)$. Specific intensity is defined as
($c$ times) the incident energy density per unit frequency interval per unit solid angle,
or, equivalently, specific intensity is the (normally incident) flux per unit solid angle.
Following the convention commonly used in SGWB analysis,
if the incident GW energy density is expressed in the units of
critical energy density for a flat universe $\rho_c := 3c^2H_0^2/(8\pi G)$,
where $H_0$ is the Hubble constant at the current epoch and
$G$ is the Newton's gravitational constant,
it can be shown that~\cite{SanjitThesis}:
\begin{equation}
I_\text{GW}(f,\Om) \ = \ \frac{4 \pi^2 c}{3 H_0^2} \, f^2 \, H(f) \left[ \P_+(\Om) \, + \, \P_\times(\Om) \right].
\end{equation} 
Therefore, $\P_A(\Om)$ can be interpreted as the specific intensity of SGWB for the corresponding polarization (up to a certain proportionality constant).

In general, we cannot separate $H(f)$ from
$\P_A(\Om)$, because the frequency power spectrum $H(f)$ could also
depend on the direction $\Om$. A more general treatment would use a
quantity like $\P_A(\Om,f)$ which describes both frequency and angular
distribution of SGWB power together. But for a small enough bandwidth
the assumption may be justified, as the signal is expected to have a
smooth profile of the power spectra.
Further, it should be noted that $\P_A$ is actually a second rank
tensor and should be represented by two indices as $\P_{AA'}$, but
because we assume that the two polarizations are uncorrelated the
quantity $\P_{+ \times}$ is zero and thus absent. To avoid unnecessary
indices and with a slight abuse of notation, we therefore write $\P_A$
with a single index instead of $\P_{AA'}$.

We consider two GW detectors located at $\x_I(t), I = 1,2$. The
detector frames are denoted by the coordinates $(X_I, Y_I, Z_I)$ where 
the arms of the
respective detectors lie along the $(X_I, Y_I)$ axes. Then the
detector tensors $d_I$ are given by: 
\be d_I = \h \left[
\mathbf{\hat{X}}_I \otimes \mathbf{\hat{X}}_I \ - \ \mathbf{\hat{Y}}_I
\otimes \mathbf{\hat{Y}}_I \right] \, .
\ee 
The factor $\h$ is inherited from the geodesic deviation equation. Owing 
to the Earth's rotation, the detector coordinates and $d_I$ are functions 
of $t$. Thus in matrix form: \be d_I (t) = \R^T (t)
\cdot d_I (t = 0) \cdot \R (t) \,, \ee where $\R (t)$ is the rotation
matrix describing a rotation of angle $\omega_E t$ around the Earth's
rotation axis, namely, the $z$-axis. Here $\omega_E$ is the Earth's
sidereal angular velocity $\approx 7.3 \times 10^{-5}$ radians/sec.

The strain $h_I (t)$ in the $I^{th}$ detector is given by: 
\be
h_I (t) = h_{ij} (t, \x_I (t)) d_I^{ij} (t).  \ee 
We define the antenna pattern functions as
\be F_I^A(t;\Om) =
e^A_{ij} (\Om) \, d_I^{ij} (t)
\ee
for a wave incident from the direction $\Om$.

Contracting Eq. (\ref{eq:plwave})
with the detector tensors $d_I$, the signal amplitudes can be
expressed in terms of the antenna pattern functions as:
\begin{equation}
h_{I}(t) = \int_{-\infty}^\infty \d f \int_{S^2} \d \Om \, \ft{h}_A(f,\Om)  F^A_{I}(t; \Om) \, e^{2\pi i f (t + \Om\cdot\mathbf{x}/c)} .
\label{eq:sigampl}
\end{equation}
The baseline $\Dx$ at any time $t$ (in matrix form) is given by:
\be
\Dx (t) = \R^T \cdot \Dx (t=0) \,.
\ee

The detector output $s_I (t)$ is a sum of the GW signal $h_I(t)$ and noise $n_I(t)$
\be
s_I(t) = h_I(t) \ + \ n_I(t) \,.
\ee
Statistically, the gravitational-wave strains $h_I (t)$ are uncorrelated 
with the detector noise $n_J (t)$; that is, the following four correlations,
in the time domain are zero:
\be
\langle h_I (t) \, n_J(\tp) \rangle \ = 0 ~~~I, J = 1,2 \,, \label{eq:tCorrZero}
\ee
where $t, \tp$ are any two time instants. We also assume that the
noise in different detectors is uncorrelated; that is, $\langle
n_1 (t) \, n_2 (\tp) \rangle \ = 0$. This assumption is not
unreasonable when the detectors are widely separated. Thus the only
possible correlation is between $h_1 (t)$ and $h_2 (t')$. 
The statistic that we construct in the next subsection is based
on this fact.

\subsection{The Cross-correlation Statistic for the Directed Search}

Stochastic GW signals inherently can arrive from any direction with
any amplitude. Moreover, they are characterized by the statistical
expectation values of energy density. The noise in the two 
detectors are assumed to be essentially independent. In this
situation the cross-correlation of the data from the two detectors 
is an appropriate statistic for
detecting and observing stochastic GW. In order to optimize the SNR,
the cross-correlation statistic for the directed search involves a
direction-dependent filter function $Q(t, \Om ;t',t'')$, which
connects sidereal time $t'$ of one detector's data to $t''$ of the
other detector's data to match the phases of the GW strains in the
detectors. The filter does a inverse noise weighting using the PSDs of
the two detectors in order to suppress noisy frequency bands
and enhance the SNR by assigning relatively large weight to the sensitive
regimes of the detectors and the bands where the source is expected to emit more power.
(In general, $Q$ will depend on $\P_A (\Om)$. However, for 
the directed search that we envisage here, as we will see later, the
$\P_A(\Om)$ are delta functions and, therefore, the filter $Q$ only depends 
on $\Om$.) 

The sampling interval with which the data are sampled is determined by the Nyquist frequency 
of the stochastic signals of interest, and can be well below a millisecond, corresponding to several kilohertz. While it is possible to compute the filter function on a time segment of this  sampling interval, it is erroneous to do so for the physical problem at hand. This is because
the signals at the two detectors will be incoherent on time-scales smaller
than the light-travel time,  $t_d$, of that baseline, which is at most a few 
tens of milli-seconds for ground-based detectors. Thus
$t_d$ sets the lower limit on the duration of the time segment
on which the filter function should be computed. The upper limit, $\tau$, is 
set by the smaller of the timescales, on which the data are stationary and the timescale on which
the detector orientation changes appreciably. We thus divide the data into time segments, $\Delta t$, such that $t_d \ll \Delta t \ll \tau$. The time segments used currently in LSC data analysis vary from $32$ to $192$ seconds, and are consistent with these limits. 

The final statistic for the full observation time $T$ is obtained by linearly combining the cross-correlations over the smaller time-intervals as a weighted
sum. The filter is optimized for each time segment, $I_k = [t_k -\Dt/2,
t_k + \Dt/2]$, at sidereal time $t_k$ and the statistic for the $k^{th}$
segment is given by:
\begin{eqnarray}
&& \Delta S (t = t_k, \Om) ~ \equiv ~  \Delta S_k (\Om) \nonumber ~~~~~~~ \\
&=& \int_{I_k} \d t' \int_{I_k} \d t'' \, s_1(t') \, s_2(t'') \, Q(t_k, \Om; t',t'') \,. ~~~~~~~ 
\label{eq:defDS} 
\end{eqnarray}

The final cross-correlation statistic $S$ for all the $n=T/\Dt$
sidereal time bins can then be obtained by combining the $\Delta S_k$
as a weighted sum as follows:
\be
S (\Om) = \sum_{k=1}^n w_k(\Om) \, \Delta S_k  \,,
\ee
where, the $n$ quantities $w_k (\Om )$ are to be chosen so that the
SNR for the statistic $S (\Om )$ 
is maximized. We denote the SNR of $S$ by $\rho_S = \mu_S / \sigma_S$,
where $\mu_S$ and $\sigma_S$ are the mean and standard deviation of
$S$ respectively. We use normalized weights $\sum_{k=1}^n
w_k =1$.

The $\mu_S, \sigma_S$ and the SNR $\rho_S$ are given in terms of
the means $\mu_k = \langle \Delta S_k \rangle$ and variances
$\sigma_k^2 = \langle \Delta S_k^2 \rangle - \langle \Delta S_k
\rangle^2$ of the individual {\em mutually uncorrelated} time segments as follows:
\bea
\mu_S &=&  \sum_{k=1}^n w_k \, \mu_k \,, \nonumber \\
\sigma_S^2 &=& \sum_{k=1}^n w_k^2 \, \sigma_k^2 \,, \nonumber \\
\rho_S &=& \left[ \sum_{k=1}^n w_k \, \mu_k \right] \bigg\slash \left[\sum_{k=1}^n w_k^2 \, \sigma_k^2 \right ]^{\h} \, . \label{eq:combinedStat}
\eea
The maximization is achieved elegantly by invoking the Schwarz
inequality.  We consider an $n$-dimensional Euclidean space equipped
with the usual scalar product in which $\vec{w}$ is a vector with
components $w_k$. We define $\vec{\kappa}$ and $\vec{\rho}$ having the
components $\kappa_k = w_k \sigma_k$ and $\rho_k = \mu_k / \sigma_k$
respectively. Then, we may write $\rho_S \equiv \hat {\mathbf \kappa}
\cdot \vec{\rho}$ where the vector $\hat {\mathbf \kappa}$ is a unit
vector in the direction of $\vec{\kappa}$. $\rho_S$ is maximized when
$\vec{\kappa}$ points in the direction of $\vec{\rho}$. Thus, $w_k
\propto \mu_k / \sigma_k^2$, the proportionality factor being $\left[
\sum_{k=1}^n \mu_k \, \sigma_k^{-2} \right]^{-1}$. We have the
results:
\bea
S &=& \frac {\sum_{k=1}^n \mu_k \, \sigma_k^{-2}\, \Delta S_k}{\sum_{k=1}^n \mu_k \, \sigma_k^{-2}} \,, \\
\rho_S &=& |\vec{\rho}| = \left \{ \sum_{k=1}^n \mu_k^2 / \sigma_k^2 \right \}^{\h} \,.
\eea
Further, we are free to choose the normalization of the filter $Q$. We
choose the normalization such that all $\mu_k$ are made identical,
equal to $\mu$, say. This simplifies the statistic $S$ and its SNR
$\rho_S$:
\bea
S &=& \frac {\sum_{k=1}^n \Delta S_k \sigma_k^{-2}}{\sum_{k=1}^n \sigma_k^{-2}} \,, \label{stat} \\
\rho_S &=& \mu  \left \{ \sum_{k=1}^n \sigma_k^{-2} \right \}^{\h} \,,
\eea
which are their final forms.
Normally the GW signal is expected to be weak so that even after its
integration over $\Dt$ it is still smaller than the noise, that is, $|
\mu_k | \ll \sigma_k$. Therefore the variance in each segment obeys
$\sigma_k^2 \approx \langle \Delta S_k^2 \rangle $. The signal however
builds up when we integrate over all the time segments during the
observation time. Secondly, from Eq. (\ref{stat}), it is evident that
noisy time intervals contribute less to the statistic $S$ leading to 
its optimal character.

\section{The Convolution Equation for the Sky Map}
\label{sec:convEq}

Since the GW power spectra and the detector noise are modeled in the
frequency domain, it is convenient to formulate the whole analysis in 
that domain. However, since the detector output is a time-series, it is
pertinent to ask over what time-duration must one compute their Fourier
transforms. As discussed in the last section, 
typical acceptable segment sizes are a few tens to a few hundreds of seconds. 
Fourier transforms computed over such ``small'' (vis \`a vis the total 
observation) time scales are termed as {\it short-term Fourier
transforms (SFTs)}. Each SFT becomes a function of time $t$ as well
because $t$ is essentially the identifier of the segment. Thus we have a
time-frequency representation of the data. As we shall see, this
representation is most suitable for further analysis and has also been
used in previous literature~\cite{alnrmn,ottaln,cornish}.

\subsection{Time-Frequency Analysis of the Signal and Noise}
\label{TimeFreqData}

The (approximate) SFT of a segment of detector output can be defined as~\cite{ottaln},
\begin{equation}
 \ft{s}_I(t;f) \ := \ \int_{t-\Dt/2}^{t+\Dt/2} \d t' \, s_I(t') \, e^{-2\pi i f t'} \,. 
\label{eq:defChunkFT}
\end{equation}
We retain here the convention of using `tilde' over a symbol to denote Fourier 
transform - the distinction should be evident from the context. 
Most importantly, by taking the inverse Fourier transform,
\begin{widetext}
\be
\int_{-\infty}^\infty \d f \,  \ft{s}_I(t;f) \, e^{2\pi i f t}  
= \int_{-\infty}^\infty \d f \, e^{2\pi i f t} \int_{t-\Dt/2}^{t+\Dt/2} \d t' \, s_I(t') \, e^{-2\pi i f t'} = s_I(t) \,,
\ee
\end{widetext}
the exact time series segment $s_I(t)$ can be recovered. The same
notation will be used for several other quantities as well in this
analysis.  

While constructing the statistic, it was assumed that the true 
GW strains $h_I (t)$ in the detectors are correlated, but that the 
noise streams are uncorrelated. The expression for the correlation between 
the GW strains in two detectors will be derived here, which is necessary 
for the derivation of the optimal filter in the next subsection. The
detector parameters can be approximated to be stationary over the
period of the time segment. Hence we may consider the quantities $\mathbf{d}_I$
and $\mathbf{x}_I$ to be nearly constant over a time segment and regard them
as functions of the time $t$ labeling the time segment. The SFT of the GW
strain in detector $I$ over a time segment is given by:
\begin{widetext}
\begin{eqnarray}
\ft{h}_I(t;f) &=& \int_{t-{\Dt \over 2}}^{t+{\Dt \over2}} \d t' \int_{-\infty}^\infty \d f' \int_{S^2} 
\d\Om \ft{h}_A(f',\Om) F_I^A(\Om,t) e^{2\pi i \left[f't' - ft' + f'{\Om\cdot\mathbf{x}_I(t)\over c} \right]} \, \nonumber \\
&=& \int_{S^2} \d\Om \, F_I^A(\Om,t) \int_{-\infty}^\infty \d f' \, \ft{h}_A(f',\Om) \, e^{2\pi i 
\left[(f' - f)t + f'{\Om\cdot\mathbf{x}_I(t)\over c}\right]}  \, \delta_{\Dt}(f - f') \,, 
\label{eq:ftGWBfull}
\end{eqnarray}
\end{widetext}
where summation over $A$ is implied. The $\delta_{\Dt}(f)$ is the
finite time delta function ($\snc$ function) defined by,
\begin{equation}
\delta_{\Dt}(f) \ := \ \int_{-\Dt/2}^{\Dt/2} \d t \, e^{-2\pi i f t}  = \frac{\sin{\pi f \Dt}}{\pi f}.
\end{equation}
The finite time delta function $\delta_{\Dt}(f)$ behaves as the Dirac delta function $\delta(f)$ in the limit $\Dt \rightarrow \infty$, but has the property $\delta_{\Dt}(0) = \Dt$. Hence for a large time segment $\Dt$ the SFT from detector $I$ takes the simple form:
\begin{equation}
\ft{h}_I(t,f) \ = \ \int_{S^2} \d\Om \, \ft{h}_A(f,\Om) \, F_I^A(\Om,t) \, e^{2\pi i f {\Om\cdot\mathbf{x}_I(t)/c}} \,, 
\label{eq:ftGWB}
\end{equation}

The important result of this subsection is the expectation of cross-correlation between the SFTs of time segments of detector outputs at time $t$ from the two detectors 1 and 2, which can be obtained from Eq.~(\ref{eq:ftGWBfull}) and Eq.~(\ref{eq:indGWBmodes}) as:
\begin{widetext} 
\bea
\langle \ft{h}^*_1(t,f) \, \ft{h}_2(t,f') \rangle &=&  e^{2\pi i t (f-f')} \int_{-\infty}^\infty \d f'' \, H(f'') \, \gamma(t, f''; \dom, d\P_A) \delta_{\Dt}(f''-f) \, \delta_{\Dt}(f''-f') \,,\label{eq:corrGWB} \\
\gamma(t, f; \dom, d\P_A) &=& \int_{\dom = S^2} \d\Om \left[ F_1^+(\Om,t) \, F_2^+(\Om,t) \, \P_+(\Om) \ + \ F_1^\times(\Om,t) \, F_2^\times(\Om,t) \, \P_{\times}(\Om)\right] e^{2\pi i f \Om\cdot\Dx(t)/c} \,.~~~~~~~ 
\label{eq:gamma}
\eea
\end{widetext}
The general overlap reduction function defined by Eq.~(\ref{eq:gamma}) is a generalization
of the usual overlap reduction function for the isotropic SGWB case,
first constructed by Nelson Christensen~\cite{crsfrm} and formally written
in a closed form by Flanagan~\cite{flan}.
In this case it is a more complex object. Besides the frequency,
it is also a function of the segment time $t$.  It also depends on
$\P_A(\Om)$ which is integrated over the full sky $S^2$. Thus it is a
functional of $\P_A(\Om)$. In general, when we construct our directed
filters, the integral will be restricted to a small patch of the sky,
$\dom \subset S^2$. (In principle, $\dom$ can be any (measurable) subset
of $S^2$.) Accordingly, $\P_A(\Om)$ plays the part of a weight function over
the sky and we may define the measures: \be d\P_A = \P_A (\Om) d \Om
\,.  \ee Thus, the $\gamma$ in general becomes a functional of the SGWB
power in both polarizations coming from the patch $\dom$. We therefore
separate the function arguments, $t$ and $f$ from the non-function arguments,
$\dom$ and $\P_A$, by a semi-colon. Finally, the exponential term in 
Eq. (\ref{eq:corrGWB}) before the integral is just the time-shift term in 
the Fourier transform of the segment at time $t$.

In the limit of a large time segment, Eq. (\ref{eq:corrGWB})
takes the simple form:
\begin{equation}
\langle \ft{h}^*_1(t,f) \, \ft{h}_2(t,f') \rangle \ = \ \delta(f-f') \, H(f) \, \gamma(t, f; S^2, d\P_A) \,. \label{eq:infcorrGWB}
\end{equation}

The advantage of expressing $\langle \ft{h}^*_1(t,f) \,
\ft{h}_2(t,f) \rangle$ by Eq.~(\ref{eq:corrGWB}) can be readily
realized if we put $f=f'$. In this case, the correlation given by Eq.~(\ref{eq:infcorrGWB})
diverges in the limit $\Dt \rightarrow \infty$. But, in practice, $\Dt$ {\em is} finite, and
hence, we expect a finite value for the correlation. Eq.~(\ref{eq:corrGWB}) lets us
compute that finite value of $\langle \ft{h}^*_1(t,f) \, \ft{h}_2(t,f)
\rangle$ at $f=f'$. We use the large $\Dt$ limit and replace one of
the finite time delta functions $\delta_{\Dt}(f''-f)$ in the integrand
of Eq.~(\ref{eq:corrGWB}) by the Dirac delta function $\delta(f''-f)$,
while treating the other $\delta_{\Dt}(f''-f)$ as a normal function
and put $\delta_{\Dt}(0) = \Dt$. We get,
\begin{equation}
\langle \ft{h}^*_1(t,f) \, \ft{h}_2(t,f) \rangle \ = \ \Dt \, H(f) \, \gamma(t, f; S^2, d\P_A).
\label{h1h2}
\end{equation}
This formula was derived by following the same procedure as prescribed in \cite{alnrmn},
so, not surprisingly, for isotropic backgrounds our result matches the formula obtained in \cite{alnrmn}. This result is important for injecting test signals in the detector output~\cite{Bose:2003nb}.
 
Next we describe the properties of detector noise in a finite
time segment.
 
The noise in the segment labeled by $t$ is a time series $n_I(t)$
in a detector $I$.  Its SFT is given by:
\begin{equation}
\ft{n}_I(t;f) \ := \ \int_{t-\Dt/2}^{t+\Dt/2} \d t \, n_I(t) \, e^{-2\pi i f t}.
\end{equation}
We take the mean to be zero: $\langle n_I(t) \rangle = \langle
\ft{n}_I(t;f) \rangle = 0$. Since $n_I(t)$ is real, its SFT
obeys the reality condition, $\ft{n}_I^*(t;f) \
= \ \ft{n}_I(t;-f)$. The noise in a detector is uncorrelated with the
noise in another detector and with the GW signal,
i.e., $\langle n_1(t) \, n_2(t') \rangle = \langle h_1(t) \, n_2(t') \rangle =
\langle n_1(t) \, h_2(t') \rangle = 0$ [see Eq.~(\ref{eq:tCorrZero})]. These relations also hold for
their corresponding SFTs. The length of the time segment is usually kept a few tens of
seconds long, over which the detector noise can be regarded as stationary. Thus
$\langle n_I(t') \, n_I(t'')\rangle$ is a function of $t''-t'$,
provided both $t',t''$ are in the same segment centered at time
$t$. Then, using the fact that $n_I(t)$ is real, we have,
\begin{equation}
\langle n_I(t') \, n_I(t'')\rangle \ = \ \frac{1}{2} \int_{-\infty}^{\infty} \d f \, P_I(t;|f|) \, e^{2\pi i f (t''-t')},
\end{equation}
where $P_I(t;f)$ is the {\em one-sided} noise PSD. This noise PSD is
also a function of time $t$ as detector noise is non-stationary.
The correlation between the corresponding SFTs can be easily
obtained from the above relations:
\begin{widetext}
\be
\langle \ft{n}_I^*(t;f) \, \ft{n}_I(t;f') \rangle =  {1 \over 2} \int_{-\infty}^{\infty} \d f''  P_I(t;|f''|) \delta_{\Dt}(f''-f) \, \delta_{\Dt}(f''-f') \,. 
\label{eq:chunkNoisePSD}
\ee
\end{widetext}
In the limit of large length of the time segment, we arrive at the usual formula
\begin{equation}
\langle \ft{n}_I^*(t;f) \, \ft{n}_I(t;f') \rangle \ = \ {1 \over 2} \, \delta(f-f') \, P_I(t;|f|). \label{eq:GWnoisecorr}
\end{equation}
Again, the advantage of using Eq.~(\ref{eq:chunkNoisePSD}) in expressing 
$\langle \ft{n}_I(t;f) \,\ft{n}_I(t;f') \rangle$ becomes evident when 
we set $f=f'$. The usual formula,
Eq.~(\ref{eq:GWnoisecorr}), involving Dirac delta function diverges, whereas,
in practice, that expression is actually finite because of a finite observation
time. However, in Eq.~(\ref{eq:chunkNoisePSD}), if we replace one
finite time delta function by the Dirac delta function and treat the
other as a normal function we obtain a finite result,
\begin{equation}
\langle | \ft{n}_I(t;f) |^2 \rangle \ = \ {1 \over 2} \, \Dt \, P_I(t;|f|) \,.
\label{eq:GWChunkNoise}
\end{equation}
This is used in our work when generating simulated noise for our test analyses.

\subsection{Optimal Filters for Anisotropic Searches and the Directed Search}

The aim of this subsection is to construct an optimal filter to
maximize the SNR of the cross-correlation statistic over the small
time-segments. We essentially generalize the analysis presented in \cite{alnrmn} 
for isotropic background search to the case of anisotropic background search.

The optimal filter depends on the theoretical model of
the SGWB, that is, on $P_A(\Om)$ and $H(f)$. First we will derive the
filter for the general case of the anisotropic search and then
specialize to the directed search. Our first goal is to compute the
SNR $\rho (t)$ of the statistic $\Delta S (t)$ over the time segments of
length $\Dt$ at time $t$.

The equation for the general case is the generalization of Eq. (\ref{eq:defDS}):
\be
\Delta S(t) =
\int_{I(t)} \d t' \int_{I(t)} \d t'' \, s_1(t') \, s_2(t'') \, Q(t;
t',t'') \,,
\label{defgen}
\ee
where $I(t)$ is the interval $[t - \Dt/2, t + \Dt/2]$. Here we have suppressed the model dependence of $Q$.  Assuming that the noise in both detectors and the earth are stationary within the duration of each time segment, we may write $Q(t;t',t'') = Q(t;t'-t'')$, which allows us to expand the filter in terms of its SFT $\ft{Q}(t,f)$ as,
\begin{equation}
Q(t;t'-t'') \ = \ \int_{-\infty}^\infty \d f \, e^{2\pi i f(t'-t'')} \, \ft{Q}(t,f).
\end{equation}
Thus in terms of SFTs the statistic can be expressed as,
\begin{equation}
\Delta S(t) \ = \ \int_{-\infty}^\infty \d f \, \ft{s}_1^*(t;f) \, \ft{s}_2(t;f) \, \ft{Q}(t,f). 
\label{eq:ccGWB}
\end{equation}
The mean of the statistic $\Delta S(t)$ is:
\be
\mu (t) \ := \ \langle \Delta S(t) \rangle \ = \ \int_{-\infty}^\infty \d f \, \langle \ft{s}_1^*(t;f) \, \ft{s}_2(t;f) \rangle \, \ft{Q}(t,f) \,.
\ee
Replacing $s_I$ by $h_I$ within the ensemble average, since the relevant correlations of signal and noise are zero except for $\langle \ft{h}_1^*(t;f) \, \ft{h}_2(t;f) \rangle$, we obtain from 
Eq. (\ref{h1h2}),
\be
 \mu (t) = \Dt \int_{-\infty}^{\infty} \d f \, H(f) \, \gamma(t, f; S^2, d\P_A) \, \ft{Q}(t,f) \,.
\label{eq:meanDeltaS}
\ee
Here the $H$ and $\P_A$ are the actual quantities pertaining to the SGWB source. The corresponding quantities of the theoretical model are hidden inside the filter $Q$.

The variance after a routine but fairly involved calculation is obtained as:
\bea
\sigma^2 (t) &=& \langle \left[ \Delta S(t) - \langle \Delta S(t) \rangle \right]^2 \rangle \nonumber \\
&= &  {\Dt \over 4} \int_{-\infty}^\infty \d f \, P_1(t;|f|) \, P_2(t;|f|) \, |\ft{Q}(t,f)|^2. ~~~~~~~ 
\label{eq:varDeltaS}
\eea 
From these results the SNR $\rho (t)$ for the segment at time $t$
can be computed. However, we need to maximize the SNR over each time
segment. Here again we follow the prescription presented in \cite{alnrmn} -
we invoke the Schwarz inequality. To this
end it is convenient to define a scalar product of two functions $A$
and $B$ on each time segment, labeled by $t$, as,
\begin{equation}
(A,B) (t) := \int_{-\infty}^{\infty} \d f P_1(t;|f|) \, P_2(t;|f|) \, \ft{A}^*(t;f) \, \ft{B}(t;f) \,. \label{eq:scProdGWB}
\end{equation}
The norm of a function $A (t)$ is defined as $\nrm A \nrm^2 = (A, A)$. Then the mean and variance are given in terms of the scalar product as follows:
\bea
\mu (t) &=& \Dt \left ( \frac {H(f) \gamma(t, f; S^2, d\P_A)}{P_1(t;|f|) \, P_2(t;|f|)}, Q (t)\right ) \,, \label{eq:muGWB} \\
\sigma^2 (t) &=& \frac{1}{4} \Dt \nrm Q \nrm^2 (t) \,.\label{eq:sigSqGWB}
\eea
The SNR $\rho (t)$ is just the ratio $\mu (t) / \sigma (t)$. The SNR is maximized when the ``signal'' and the ``filter'' vectors are parallel, which happens when,
\begin{equation}
\ft{Q} (t,f)  = \lambda (t) \, \frac{H(f) \, \gamma^*(t, f; S^2, d\P_A)}{P_1(t;|f|) \, P_2(t;|f|)}, 
\label{eq:filterGWB}
\end{equation}
where $\lambda (t)$ is a (real) proportionality constant for the time
segment. It is a function of the segment time $t$. This function will be
chosen so that the SNR of $S$ for the full observation is
maximized.

For the optimal filter, the expression for
the mean given by Eq.~(\ref{eq:meanDeltaS}) simplifies to:
\be
\mu (t) = \frac{\Dt \nrm Q \nrm^2
(t)}{\lambda (t)}.
\ee
We exploit the freedom of choosing $\lambda
(t)$ by setting $\mu (t) = 1$ for each time segment. This immediately
gives: \bea \lambda (t) &=& \Dt \nrm Q \nrm^2 (t) \,, \\ \sigma^2 (t)
&=& \frac{\lambda (t)}{4} \,.  \eea We further require to find
$\lambda (t)$ explicitly. This is done by computing $\nrm Q
\nrm^2$. We find $\nrm Q \nrm^2 = \lambda^2 P_{NW}^2$ where, \be
P_{NW}^2 (t) = \int_{-\infty}^{\infty} \d f \frac{H^2(f) \, |\gamma(t,
f; S^2, d\P_A)|^2}{P_1(t;|f|) \, P_2(t;|f|)} \,.
\label{pwr2}
\ee
The integral on the right-hand side of the above equation is positive definite: Its integrand contains the fourth power of the GW amplitude. Therefore, it may be denoted by the square of a real quantity $P_{NW}$, which is determined by the GW power accessible to the network of the two detectors. The above equation gives the normalizations:
\bea
\nrm Q(t) \nrm^2 &=& [ \Dt P_{NW} (t) ]^{-2} \,, \nonumber \\
\lambda (t) &=& [\Dt P_{NW}^2 (t) ]^{-1} \equiv 4 \sigma^2 (t) \,.
\label{var}
\eea
The optimal statistic is then easily obtained as in Eq.~(\ref{stat}). We replace the sum in that equation by an integral over the segment time $t$:
\be 
S = \frac{\int \Delta S (t) \sigma^{-2} (t) \d t}{\int \sigma^{-2} (t) \d t} \,,
\label{Fstat}
\ee 
where the integration is over the observation time (which could
consist of disconnected time intervals).   

The filter given by Eq.~(\ref{eq:filterGWB}) is {\em the general optimal filter} to search for any anisotropic SGWB, which, unfortunately, requires the knowledge of $H(f)$ and $\P_A(\Om)$. In practice, we do not have an exact {\it a priori} model for
$H(f)$ and $\P_A(\Om)$, which anyway we are trying to measure! We then
must use models for those quantities, $H'(f)$ and $\P'_A(\Om)$, to search for different
anisotropic backgrounds. These models will be used to construct the
model-dependent overlap reduction function $\gamma(t, f; \dom,
d\P'_A)$ and the filter $Q$.

The angular power distribution
for only one unit point source on the sky in the direction $\Om$ with
equal power in both the polarizations can be expressed as:
\begin{equation}
\P_A(\Om') \ = \ \delta(\Om' - \Om).
\end{equation}
This immediately simplifies the expression for the overlap reduction function because now the integral in Eq. (\ref{eq:gamma}) simplifies owing to the delta-function, and $\gamma$ becomes a function of $\Om$ as well. Therefore, we have:
\bea
\gamma(\Om, t, f) &=& \Gamma (\Om, t) e^{2\pi i f \Om\cdot\Dx(t)/c} \,, \label{eq:dirgamma} \\ 
\Gamma (\Om, t) &=& \sum_{A} F_1^A(\Om,t) \, F_2^A(\Om,t) \,.
\label{Gmm}
\eea
Unlike the time-independent overlap reduction function of the isotropic SGWB case, the {\em direction-dependent overlap reduction function}, $\gamma ( \Om, t, f)$, accepts power from {\em all} the frequencies and in fact has infinite bandwidth in the limit of vanishing pixel area. So the bandwidth of the filter $Q$ in this case would only be limited by the bandwidth of the detectors through the coefficients $P_I(t;f)$.

If, instead, the source has a finite spatial extent, the bandwidth would be limited, because the integral in Eq.~(\ref{eq:gamma}) would have to be performed over the solid angle $\dom$ subtended by the source. If one takes a small patch of the sky of size $(\Delta \theta, \Delta \phi )$ around some fixed (source) direction $\Om_0 := (\theta_0, \phi_0)$, it is easy to show that $\gamma$ is essentially a product of sinc functions in $(\Delta \theta, \Delta \phi)$. In fact:
\bea
&& \int_{\dom} \d \Om \, e^{2 \pi i f (\Om - \Om_0). \Dx/c} \nonumber \\
&& \qquad = \ |\Delta \Omega| \, \snc \frac{\pi f \Delta \theta \Dx. \hat{\mathbf{e_{\theta}}}}{c} \,  
\snc \frac{\pi f \Delta \phi \Dx. \hat{\mathbf{e_{\phi}}}}{c} \,, \qquad \ \,
\eea
where $| \dom | = \sin \theta \Delta \theta \Delta \phi$ is the solid angle subtended by the patch $\dom$. In the integral, the factor $\Gamma (\Om, t)$ remains nearly constant. The $\snc$ functions go to zero when their arguments reach $\pi$ radians. Taking this definition as the bandwidth and taking $| \Dx |/c \sim 10$ms for the two LIGO detectors, the bandwidth is about $750$~Hz for a square source of side $10^\circ$ on the sky.  

If there were known models for the anisotropic SGWB, the optimal filter for the general anisotropic case would have included $\P'_A(\Om)$ and we would perform a {\em full sky search for an anisotropic background}. However, no reasonable model for the anisotropic SGWB sky exists in literature and, so, blind estimations are currently the only possible alternatives.

Directed search is one blind estimation approach, where the strength of each point (pixel) of the sky is `observed' using a direction dependent filter, assuming that the other points on the sky do not contribute towards the observed value. So, in the directed case $Q$ becomes a function of $\Om$:
\begin{equation}
\ft{Q} (\Om, t, f; H) \ = \ \lambda (\Om, t) \, \frac{H(f) \, \gamma^*(\Om, t, f)}{P_1(t;|f|) \, P_2(t;|f|)},
\label{dirfilt}
\end{equation}
where $\lambda (\Om, t)$ is the normalization constant, which now varies from pixel to pixel. It is given by Eq. (\ref{var}), but with the $\gamma$ in the expression for $P_{NW}^2$ in Eq. (\ref{pwr2}) replaced by the simpler $\Gamma$ of (Eq. (\ref{Gmm})). The directed filter given in Eq. (\ref{dirfilt}) is an optimal filter if there is a single point source in the direction $\Om$ and {\em no sources elsewhere in the sky}. If there are other sources in the sky, as in a general anisotropic background, the filter becomes {\em suboptimal} as it stands.
However, we use the above filter to make a ``dirty'' map of the sky, which is a convolution of the the actual anisotropic background with the beam function and contains additive noise. We intend to extract information about the true background by the process of deconvolution. The convolution equation will be derived in the next subsection.

The working principle of the above filter is evidently similar to the earth rotation aperture synthesis often used in CMB and radio astronomy to make map of a certain portion or the whole sky. The phase lag between two detectors, separated by a distance $\Dx(t)$, in receiving a plane wavefront from a certain direction $\Om$, as shown in Figure~\ref{radiometer}, is compensated in the filter via the phase factor $\exp[{2\pi i f \Om\cdot\Dx(t)/c}]$. As the earth rotates this factor adjusts, so that, waves from the given direction are coherently added, while the waves from other directions tend to cancel out. Note that, we did not introduce the phase factor by hand, it appeared automatically through the process of the maximization of the SNR. Though the whole radiometer analysis is based on this principle, the idea is clearly realized in the directed search analysis.\\

\subsection{The Integral Equation for the Directed Search}
\label{integeqn}

In this subsection we set up the convolution equation, which is an integral equation
for the statistic $S(\Om)$ - the dirty map of directed search.
The kernel of the integral equation consists of beam functions that we define
below. The goal is to obtain the power in both polarizations $\P_A(\Om)$,
given the statistic $S(\Om)$. To this end we take the expectation
value of the statistic $S (\Om)$ in Eq.~(\ref{Fstat}) and use
Eq.~(\ref{eq:meanDeltaS}) for the expectation value of $\Delta S$
inside the integral sign. We must also use expressions for the source
overlap reduction function from Eq.~(\ref{eq:gamma}) and the directed
filter from Eq.~(\ref{dirfilt}). The final form of the (noiseless) convolution equation is given by:
\begin{widetext}
\be
s(\Om) \ \equiv \ \langle S(\Om) \rangle \ = \ \int_{S^2} \d\Omp \left[ B^+(\Om,\Om') \, \P_+ (\Om') \ + \ B^\times(\Om,\Om') \, \P_\times (\Om')\right]\,, \label{inteq}
\ee
where the beam functions $B^A (\Om, \Omp)$ are defined as:
\bea
B^A(\Om, \Omp) &=& \Lambda(\Om) \int \d t \int_{- \infty}^{\infty} \d f \frac{H(f) H'(f)}{P_1(t;|f|) \, P_2(t;|f|)} F_1^A(\Om',t) F_2^A(\Om',t) \Gamma(\Om,t) e^{-2\pi i f {\DOm \cdot \Dx(t) \over c}} \, \label{beamInteq} \\
\Lambda^{-1}(\Om) &:=& \frac{1}{\T} \int \lambda^{-1} (t, \Om) \, \d t \label{sigmaInteq}\,.
\eea
\end{widetext}
The function $H'(f)$ is the model power spectrum of the source we
insert in the kernel and $\DOm = \Om - \Omp$. We measure $\langle
S(\Om) \rangle$ and from the kernels $B^A (\Om, \Omp)$, we propose to
solve the integral equation for $\P_A (\Om)$. Physically, we may
expect the power in both polarizations to be the same, that is, $\P_+
(\Om) = \P_\times (\Om) = \P (\Om)$, say, the kernel then is just the
sum of the two individual kernels of each polarization,
\begin{widetext}
\bea
B (\Om, \Omp) &=& B^+ (\Om, \Omp) + B^\times (\Om, \Omp) \nonumber \\
&=& \Lambda(\Om) \int \d t \int_{-\infty}^\infty \d f \frac{H(f) H'(f)}{P_1(t;|f|) \, P_2(t;|f|)} \Gamma(\Om',t) \, \Gamma(\Om,t) \, e^{-2\pi i f {\DOm \cdot \Dx(t)/c}} \,.
\label{kernel}
\eea
\end{widetext}
Our numerical deconvolution strategy is described in the next section. But before we do that we examine the kernel in Eq. (\ref{kernel}) and try to understand it from a physical point of view. This will afford us some insight into the beam patterns associated with directed filters.

It is worth noting that the beam function $B (\Om, \Omp)$ is not symmetric only due to the leading normalization factor $\Lambda(\Om)$, which comes from the normalization of the statistic $S(\Om)$. We make use of this observation to introduce a symmetric kernel in section~\ref{sec:numRes}, which is advantageous for numerical deconvolution.

\subsection{The Stationary Phase Approximation of the Kernel and its Singular Value Decomposition}

The GW radiometer beams are not pointed but have a spread out profile, which varies with sky position. Thus in order to make progress towards deconvolving the GW sky map we try to understand the beam pattern.  We find that the Stationary Phase Approximation (SPA) of $B (\Om, \Omp)$, given in Eq. (\ref{kernel}), yields useful results. It is essentially the exponential term $\exp[-2\pi i f {\DOm \cdot \Dx(t)/c}]$ containing the phase that determines the integral - the integrand constructively contributes when the phase in the integral in Eq. (\ref{kernel}) is stationary. We also use the fact that rest of the functions in the integrand vary slowly with time, so that they effectively behave as constants as far as the integral is concerned. 

\begin{figure}[h]
\centering
\subfigure[~Numerical beam pattern.]{\label{fig:GWBeam-a}\includegraphics[width=0.4\textwidth]{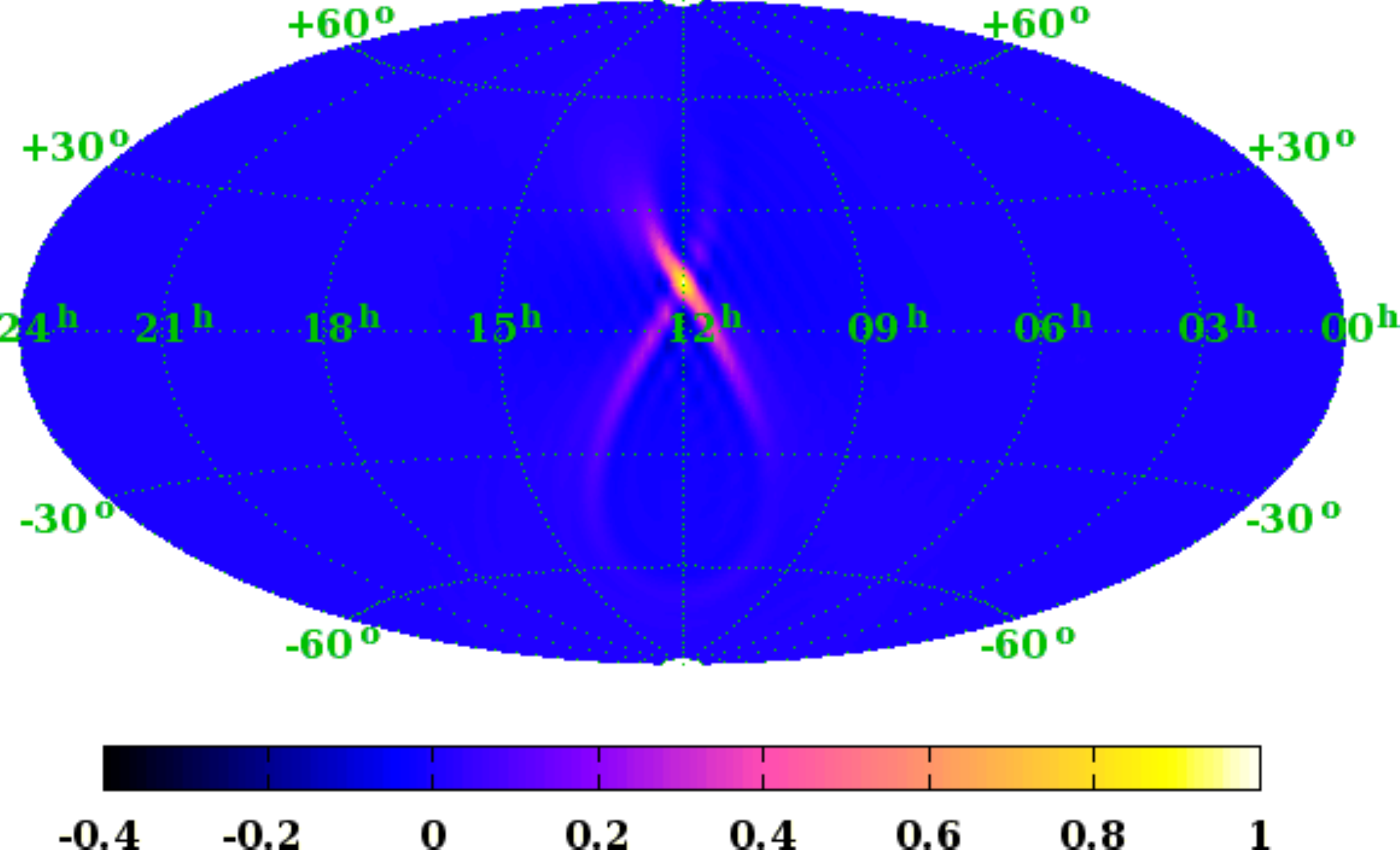}}\\
\subfigure[~Theoretical beam pattern obtained by stationary phase approximation (SPA)]{\label{fig:GWBeam-b}\includegraphics[width=0.4\textwidth]{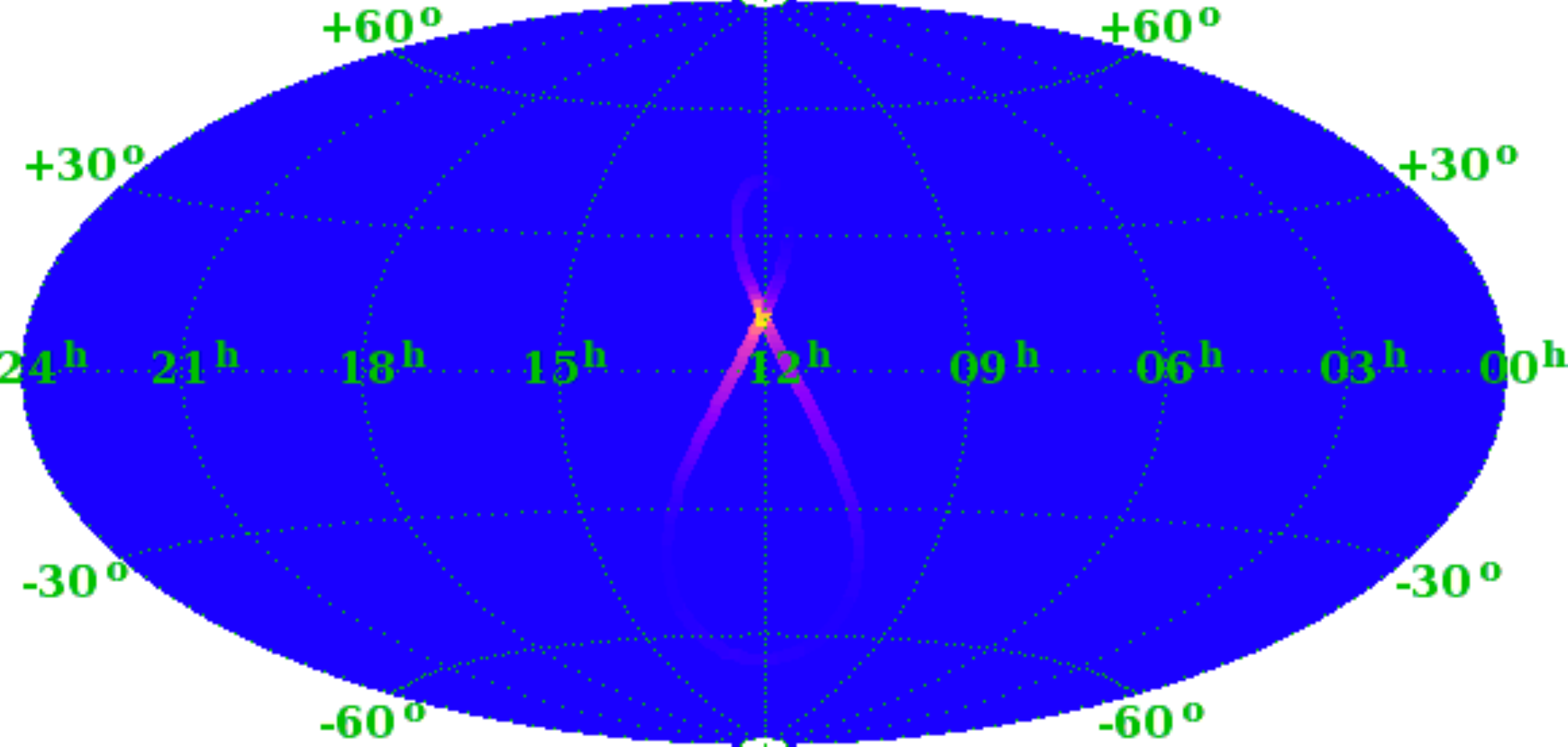}}
\caption{Illustration of the agreement between numerical and theoretical GW radiometer beam patterns at declination $+12^\circ$ for the LIGO detectors at Livingston and Hanford (with white noise, upper cut-off frequency of $1024$~Hz, $H(f)$=constant and observation time of one sidereal day). Same colormap has been used for both panels. Note that, the {\em shape} of the beam does not depend on the right ascension of the pointing direction for observation time of full sidereal day(s).}
\label{fig:GWBeam}
\end{figure}

We obtain the beam function for a unit point source at $\Om_0 \equiv (\theta_0,\phi_0)$. We  write $\DOm := \Om - \Om_0$. Note that $\DOm$ may not be necessarily small; the points $\Om, \Om_0$ can lie anywhere on the unit sphere. By performing a numerical computation for an observation time of one sidereal day, we find that for low declinations, the beam is shaped like the figure of ``8'', as shown in Figure~\ref{fig:GWBeam-a}, while as one goes higher in declination, the ``8'' smoothly turns into a ``tear drop''.

\begin{figure}[h]
\centering
\includegraphics[width=0.4\textwidth]{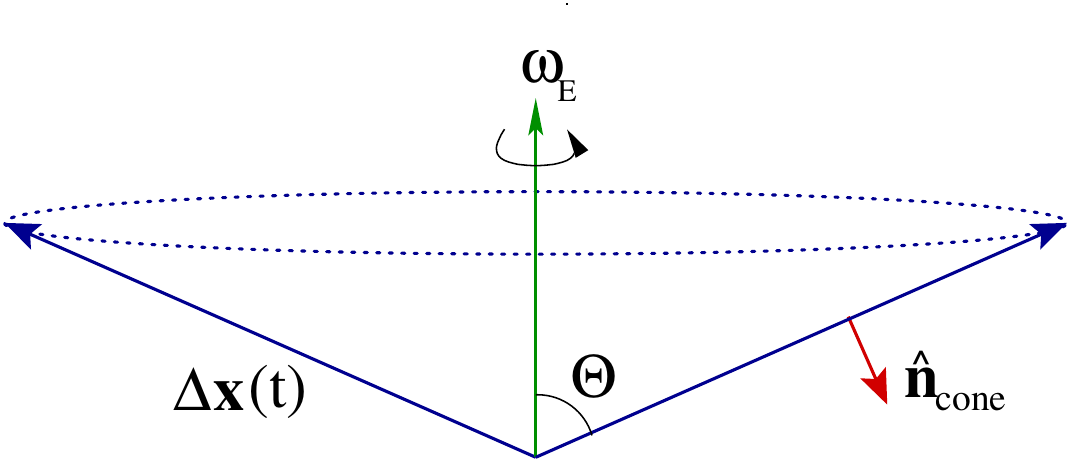}
\caption{The baseline formed by two detectors, $\Dx(t)$, traces out a cone as the earth rotates. A schematic diagram is shown here. The vector $\ncone(t)$ is normal to the baseline as well as the cone.} 
\label{fig:cone}
\end{figure}

With the application of the SPA we can explain the shape of the beam. The integrand in the kernel usually oscillates rapidly, because of the exponential phase term, except when the phase is stationary. The kernel is a double integral over $f$ and $t$ and therefore the SPA must be carried out in two dimensions. Setting the first derivative of the phase with respect to both variables $f$ and $t$ equal to zero, we obtain:
\begin{eqnarray}
\DOm \cdot \Dx(t) &=& 0, \label{eq:phase1}\\
\DOm \cdot \Ddx(t) &=& 0, \label{eq:phase2}
\end{eqnarray}
where $\Ddx(t):=\d \Dx(t) / \d t$. The detector separation vector $\Dx(t)$ rotates about the earth's rotation axis ($z$-axis in our coordinate system) with the angular velocity $\omega_E$. Geometrically $\Dx(t)$ traces out a right circular cone with $z$-axis as its symmetry axis [see Figure~\ref{fig:cone}]. It is explicitly given by:
\be
\Dx (t) =  \Delta R (\sin \Theta \cos \omega_E t, \sin \Theta \sin \omega_E t, \cos \Theta ) \,,
\ee
where $\Delta R = |\Dx (t)|$ is the constant distance between the detectors. As $\omega_E t$ ranges from 0 to $2 \pi$, $\Dx (t)$ traces out a cone with half angle $\Theta$.
The half-angle $\Theta$ of the cone, $0 \le \Theta \le \pi$, is given by,
\begin{equation}
\cos \Theta \ = \ \frac{\cos\theta_1 - \cos\theta_2}{\sqrt{2[1 - \cos\theta_1 \cos\theta_2 - \sin\theta_1\sin\theta_2\cos(\phi_1-\phi_2)]}},
\end{equation}
where $(\theta_I,\phi_I)$ are the detector coordinates.  For the LIGO pair of detectors, $\Theta \sim 27^\circ$.

From Eqs. (\ref{eq:phase1}) and (\ref{eq:phase2}) it is clear that the phase is stationary when $\Dx(t)$, $\Ddx(t)$ and $\DOm$ form an orthogonal triad. Since the unit vector normal to the baseline at any given time as well as to the cone
\be
\ncone(t) = \frac{\Dx(t) \times \Ddx (t)}{|\Dx(t) \times \Ddx (t)|} \,,
\ee
the SPA condition is satisfied when: 
\be
\DOm = \dom \, \ncone(t) \,,
\label{SPA_cond}
\ee 
where $\dom = |\DOm|$ can take \emph{both positive or negative} values. Since both $\Om_0$ and $\Om(t) = \Om_0 + \DOm(t)$ are constrained to lie on the unit sphere and thus both have unit norm, it follows from Eq.(\ref{SPA_cond}) that, the SPA solution $\Om (t)$ is a curve on the unit sphere given by~\cite{SanjitThesis},
\begin{equation}
\Om(t) \ = \ \Om_0 \ -\ 2 \, [\Om_0\cdot\ncone(t)] \, \ncone(t) . 
\label{SPASoln}
\end{equation}
The trajectory has been parameterized in terms of the sidereal time $t$.
One can even obtain an approximate analytical expression for the beam function along the SPA trajectory using standard SPA techniques as~\cite{SanjitThesis}:
\begin{eqnarray}
&&B(\Om(t),\Om_0)  \ =\ \Lambda(\Om(t)) \, \Gamma(\Om(t);t) \, \Gamma(\Om_0;t) \ \times \no\\
&& \qquad \frac{\sqrt{f_u} - \sqrt{f_l}}{\omega_E} \sqrt{\frac{8 \, c}{\left| [\hat{\mathbf{z}} \cdot \Dx(t)] \, [\hat{\mathbf{z}}\cdot(\Om(t) - \Om_0)] \right|}}.\qquad\ \, \label{eq:SPAResult}
\end{eqnarray}
As $t$ is varied over a full sidereal day, the shaded figure of ``8'' is generated through Eq.~(\ref{SPASoln}) and Eq.~(\ref{eq:SPAResult}) as shown in Figure~\ref{fig:GWBeam-b}. Clearly, SPA results match very well with the numerical beam pattern shown in Figure~\ref{fig:GWBeam-a}.

The case where Eq.~(\ref{eq:SPAResult}) does not apply (though the analysis still remains valid) is when the detectors are at the same latitude, as the normal to the baseline cone, $\ncone(t)$, always remains parallel to the $\hat{\mathbf{z}}$ axis, causing the denominator of Eq.~(\ref{eq:SPAResult}) to vanish. In this case the whole SPA trajectory shrinks to a point, which is the image of the pointing direction about the equatorial plane, $\Om = \Om_0 \,-\, 2 [\Om_0\cdot\mathbf{\hat{z}}] \mathbf{\hat{z}}$. The value of the beam function at the image point,
\begin{equation}
\int_0^T \d t \, \Gamma(\Om_0,t) \, \Gamma(\Om_0 \, -\, 2 [\Om_0\cdot\mathbf{\hat{z}}] \, \mathbf{\hat{z}},t) \bigg\slash  \int_0^T \d t \, [\Gamma(\Om_0,t)]^2 \, ,
\end{equation}
is also quite easy to compute. Therefore, a skymap produced by such a baseline will be a superposition of the (blurred) true sky and its (differently blurred) reflection about the equatorial plane.  In practice, a situation like this arises for the LHO-Virgo pair, as their latitudes are quite close, $46^\circ 27''$N and $43^\circ 37''$N respectively.

\begin{figure}[b]
\centering
\includegraphics[width=0.45\textwidth]{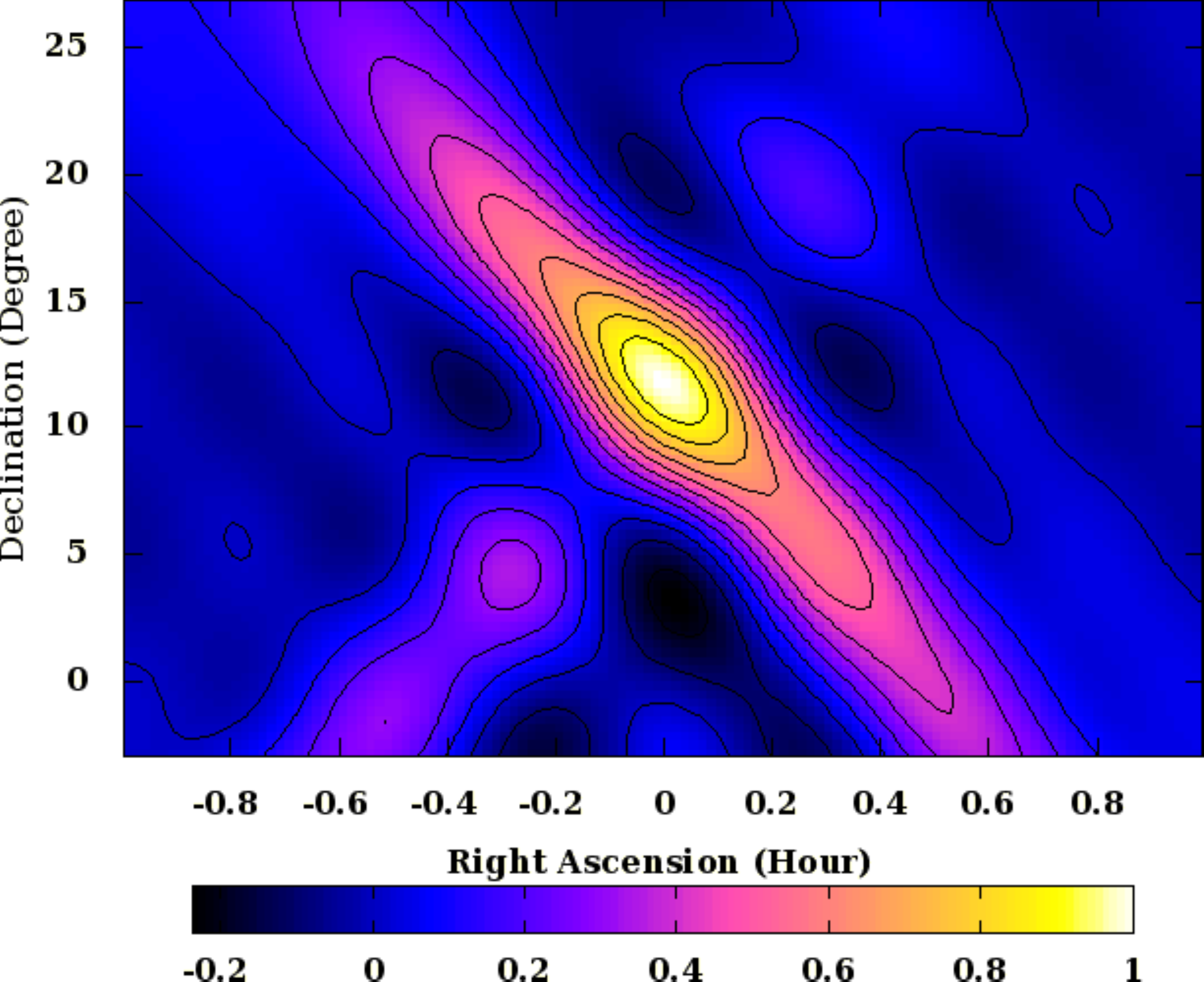}
\caption{Contour plot of GW radiometer beam pattern at declination $+12^\circ$ revealing the approximate size and the (negative) side lobes of the beam for the LIGO detectors with white noise, upper cut-off frequency $f_u = 1024$~Hz, $H(f)$=constant and observation time of one sidereal day. The contours are drawn starting from the highest level of 
$0.9$ with a difference in levels of $0.1$. The beam falls by $1/e$ in between the 5th and the 6th contour (from the highest value), which is a good indicator for the beam size. Clearly, the narrowest beam size (along the ``minor axis'') is in reasonable agreement with the theoretical prediction of $6^\circ \sim 0.1$~rad. The beam becomes broader (not shown in the figure) for real detector noise and negative source spectral index, e.g., $H(f) \propto f^{-3}$.}
\label{fig:beamContour}
\end{figure}

The SPA solution given in Eq.~(\ref{eq:SPAResult}) does not remain finite very close to the pointing direction, as the denominator vanishes. However, close to the pointing direction a better approximation is obtained by expanding the the phase term $\exp[-2\pi i f {\DOm \cdot \Dx(t)/c}]$ up to the {\em second} order, which can also be used for more accurate modeling of the core of the beam.

The stationary phase analysis also indicates an approximate resolving power of the radiometer. Since near the maximum of the beam function, the phase must not vary too much over the bandwidth, say no more than a radian, this implies that the resolving power is determined by 
$|\DOm| \sim \lambda / |\Dx|$, where the $\lambda$ corresponds to the bandwidth $f = \Delta f$ of the detectors. For the LIGO detectors, $|\Dx| \sim 3000$km. If we take the bandwidth to be $\sim 1$kHz, then $\lambda \sim 300$km, so the radiometer resolution is $\sim 0.1$ radians, that is, $\sim 6$ degrees, which we find is consistent with the numerically obtained beam profiles. A contour plot of the core of the beam of the radiometer formed by the LIGO pairs (with white noise and upper cut-off frequency of $1024$~Hz) is shown in Figure~\ref{fig:beamContour}. The plot confirms that the beam size in this case is $6^\circ \sim 0.1$~rad.

\begin{figure}[h]
\centering
\includegraphics[width=0.5\textwidth]{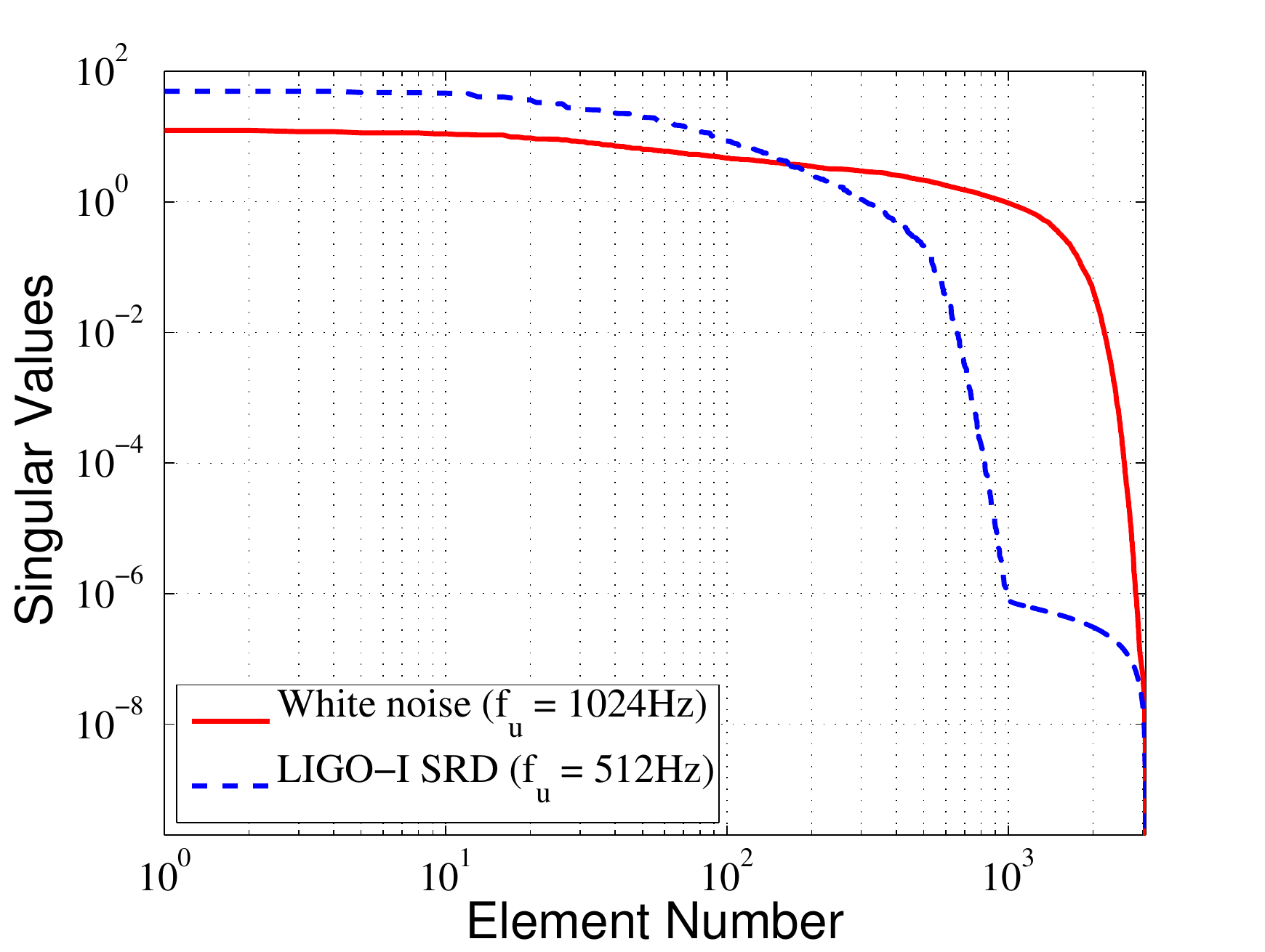}
\caption{The solid line is a plot of the singular values of the kernel for the LIGO pair of detectors with white noise, upper cut-off frequency of $f_u = 1024$~Hz and observation time of one sidereal day. From the plot it is evident that the singular values are negligible after $\sim 1000$. These results agree with the numerical and SPA results which give the size of the central spot $\sim 0.1$ radian. The dotted line shows the singular values for the same detector pair, but with LIGO-I noise PSD and upper cut-off frequency of $f_u = 512$~Hz. Clearly, the latter curve indicates a loss of angular resolution due to relatively poorer higher frequency response of the detectors.}
\label{fig:SVD}
\end{figure}

The width of the beam estimated above seen to be consistent with the `number of degrees of freedom' present in the kernel (beam). A widely used method that identifies the linearly independent modes in a linear transformation is the singular value decomposition (SVD)~\cite{golub96}. The decomposition identifies linear combinations of modes that have almost zero eigenvalues - the null subspace. It then projects out the solution orthogonal to the null subspace which spans the true degrees of freedom. The singular values of the kernel for the LIGO detectors at Hanford and Livingston for white noise with upper cut-off frequency $f_u=1024$~Hz are plotted in Figure~\ref{fig:SVD} (solid line).
The figure shows that the eigenvalues become essentially negligible after $\sim 1000$ implying that this is the number of degrees of freedom in the kernel. The numerical and the SPA analysis shows that the size of the central spot or the resolution is $\sim 0.1$ radian, which means that there are $4 \pi / (0.1)^2 \sim 1000$ independent patches in the sky. So the SVD results are consistent with the size of the beam obtained by numerical and theoretical methods.

In practice, however, the higher frequency response of the detectors is not as good, hence the achievable angular sensitivity of the radiometer becomes relatively poor. The plot of singular values for the same detector pair both having LIGO-I goal noise PSD with an upper cut-off frequency of $f_u=512$~Hz is overlaid (dashed line) on Figure~\ref{fig:SVD}. Clearly, the number of degrees of freedom, which represents the amount of information content in a map, is less in this case.
 
\section{The Maximum Likelihood Deconvolution}
\label{sec:ML}

\subsection{Unpolarised Background and Single Baseline}

We first consider the simpler case of detecting and deconvolving the signal from an unpolarised SGWB using one pair of detectors. We later indicate in the subsections that follow how this method can be extended to the more general cases of SGWBs and detector baselines.

The observed data construct, $S(\Om)$, consists of a signal and
additive noise, namely,
\be   
S(\Om) =  s(\Om) + \ n(\Om) \,.
\ee
The first term on the {\it rhs}, the expectation of the dirty map
given in Eq.~(\ref{inteq}), is a convolution of the true power in SGWB
$\P_A (\Om)$ in the two polarizations arriving from a
direction $\Om$ in the sky with corresponding beam response functions
$B^A(\Om,\Om')$. Note that, though the definitions of the above quantities
involve complex Fourier transforms, these quantities are all {\em real} owing
to the fact that they were originally derived from
real time series, so that, $\ft{s}_I^*(t;-f) = \ft{s}_I(t;f)$,
$\gamma^*(\Om, t, -f) = \gamma(\Om, t, f)$ and so on.

In this section, we construct the Maximum Likelihood (ML) estimator
for the angular power distribution $\P_A (\Om)$ given the measured
data $S(\Om)$.  For simplicity and clarity of presentation, we first
limit the analysis to the simple case where both the polarizations
follow the same angular power distribution $\P(\Om)$. This simplifies
the form of the construct to 
\begin{equation}
S(\Om) \, = \, \int_{S^2}
\d\Om' \, \big[ B_+(\Om,\Om') + B_\times(\Om,\Om') \big] \P(\Om') \, + \, n(\Om).  
\end{equation}

In practice, the sky is divided into finite number of pixels.
Then, the observed data vector is denoted by $\mathbf{S}$, whose
component $S_i := S(\Om_i)$ is the signal measured at the $i^{th}$
pixel. We similarly define the vectors
$\bm{\P}$ and $\mathbf{n}$, with components $\P(\Om_i)$ and
$n(\Om_i)$, respectively. In this notation the convolution leads to a set 
of linear algebraic equations,
\begin{equation}
\mathbf{S} \ = \ \mathbf{B} \cdot \bm{\P} \ + \ \mathbf{n} \,, \label{eq:conv}
\end{equation}
where $\mathbf{B}$ is the \emph{known}\footnote{As mentioned in the text, we assume here that we know the correct source PSD $H(f)$, which is of course {\em not} known prior to a detection. However, as long as it is known that $H(f)$ follows a power law with a finite set of possible spectral indices, one can estimate/constrain SGWB anisotropy for each possible spectral index.} beam matrix, expressed as $B_{ij} :=
B_+(\Om_i,\Om_j) + B_\times(\Om_i,\Om_j)$.
In the weak-signal approximation, the variances of the signal-noise cross terms $\ft{h}_{1,2}^*(t;f) \, \ft{n}_{2,1}(t;f)$ are much smaller than the variance of the noise-noise cross term $\ft{n}_{1}^*(t;f) \, \ft{n}_{2}(t;f)$. So the observed pixel noise is strongly dominated by the noise-noise term and can be written as,
\bea
n_i &:=& n(\Om_i) \ = \ \left[ \int \d t \, \lambda^{-1}(t,\Om_i) \right]^{-1} \ \times\\
&& \ \int \d t \, \int_{-\infty}^{\infty} \d f \, \ft{n}_1^*(t;f) \, \ft{n}_2(t;f) \, \frac{\ft{Q}(\Om_i,t,f;H)}{\lambda(t,\Om_i)} \nonumber \,;
\eea
the cross-terms $\ft{h}_{1,2}^*(t;f) \, \ft{n}_{2,1}(t;f)$ have been dropped.
The pixel noise is a sum of a large number of zero mean random numbers, where none of the addend strongly dominate (statistically) over the others. Hence, following the generalized central limit theorem~\cite{Feller}, one can argue that the pixel noise tends to be zero mean Gaussian.  If this argument is
used to calculate the variance of $n(\Om)$, we consistently get the same result as expected from Eq.~(\ref{eq:combinedStat}). After a routine but fairly involved algebra, the pixel-to-pixel noise covariance matrix, $\mathbf{N} \equiv N_{ij}$ of the dirty map turns out to be:
\begin{equation}
N_{ij} \ := \ \langle n_i \, n_j \rangle \ = \ \frac{\T}{4} \left[ \int \d t \, \lambda^{-1}(t,\Om_j) \right]^{-1}  \, B_{ij}. \label{eq:dirtyCov}
\end{equation}

The deconvolution problem is, of course, a very standard problem in many areas of
science. In particular, Eq.~(\ref{eq:conv}) is identical in structure
to the set of equations that arise in the map-making stage of CMB
experiments. The temperature anisotropy $\Delta T_i$ in a direction is
inferred from time-stream data, $d_t$ using a linear model $ d_t =
\sum_i A_{ti} \Delta T_i + n_t$.  The convolution kernel $A_{ti}$ that
relates the time domain to the pixel domain is determined by the
pointing or scan strategy as well as the beam response function of the
antenna. The noise $n_t$ is (assumed to be) Gaussian and described by
the noise covariance matrix $N_{tt'}$. As described below, a ML
solution for the sky map $\Delta T_i$ is readily obtained in this
linear model under the assumption that the noise is Gaussian.  We
adapt this technique to solve our problem since it has been applied
with great success in the CMB field and there exists an extensive
literature~\cite{cmbmap_comp} and also public domain package for
implementing it numerically~\cite{borrill99,MADCAP}. However, it should
be noted that the problem differs in two important aspects. First, in
our case there is the simplicity that the kernel connects two vectors
which are both in sky pixel space. This implies the kernel is a square
matrix for the case of single baseline and unpolarized background.
Second, in our the case the statistics of the noise is
potentially non-trivial; the noise in this situation is a complex object built by 
integrating the product of two random variables corresponding to the 
noise streams in each detector. The Gaussianity of the noise has to arise from the
generalized central limit theorem~\cite{Feller}.

We proceed assuming that the joint probability distribution of the
elements of $\mathbf{n}$ is a multivariate Gaussian
distribution~\cite{Morrison} given by the probability distribution
function:

\bea 
\mathfrak{P}(\mathbf{n}) &=& \frac{1}{(2\pi)^{\Npix/2}
\Vert\mathbf{N}\Vert^{1/2}} \, \exp{\left[-\frac{1}{2} \mathbf{n}^T
\cdot \mathbf{N}^{-1} \cdot \mathbf{n}\right]} \no\\ &=&
\frac{1}{(2\pi)^{\Npix/2}} \no\\
&&\times \exp{\left[ -\frac{1}{2}
\left( \mathbf{n}^T \cdot \mathbf{N}^{-1} \cdot \mathbf{n} +
\mathrm{Tr}[\ln{\mathbf{N}}] \right) \right]}\,,\no\\
\eea
where $\Npix$ is the total number of pixels and 
$\mathbf{N} := \langle \mathbf{n} \mathbf{n}^T \rangle$ is the 
\emph{known} noise covariance matrix [see
Eq.~(\ref{eq:dirtyCov})]. Thus, given a signal vector $\Pest$, the
probability of observing radiometer output $\mathbf{S}$ is~\cite{borrill99}, 
\bea
\mathfrak{P}(\mathbf{S}|\Pest) &=& (2\pi)^{-\Npix/2} \ \times\no\\
&&\exp\bigg[ -\frac{1}{2} \big(
(\mathbf{S} - \mathbf{B} \cdot \Pest)^T \cdot \mathbf{N}^{-1} \cdot
(\mathbf{S} - \mathbf{B} \cdot \Pest) \no\\ &&\quad\quad\quad\quad \ +\
\mathrm{Tr}[\ln{\mathbf{N}}] \big) \bigg].
\eea
Solving for $\partial\mathfrak{P}(\mathbf{S}|\Pest)/\partial\Pest=\mathbf{0}$,
using the fact that $\mathbf{N} $ is symmetric and positive definite, it is
straightforward to show that the above probability is maximum when
\begin{equation}
\Pest \ = \ \left( \mathbf{B}^T \, \mathbf{N}^{-1} \, \mathbf{B} \right)^{-1} \, \mathbf{B}^T \, \mathbf{N}^{-1} \cdot \mathbf{S},
\end{equation}
which is, therefore, the well-known result for the desired ML
estimator of the true sky map of SGWB anisotropy. The deconvolved
map will also have pixel noise, given by
\bea
\bm{n} &:=& \Pest \
- \ \bm{\P} \no\\ &=& \left( \mathbf{B}^T \, \mathbf{N}^{-1} \,
\mathbf{B} \right)^{-1} \, \mathbf{B}^T \, \mathbf{N}^{-1} \cdot
(\mathbf{B} \cdot \bm{\P} \ + \ \mathbf{n}) - \ \bm{\P} \no\\ &=& \
\left( \mathbf{B}^T \, \mathbf{N}^{-1} \, \mathbf{B} \right)^{-1} \,
\mathbf{B}^T \, \mathbf{N}^{-1} \cdot \mathbf{n}
\eea
and the pixel to pixel noise covariance matrix of the ML map is obtained as
\be
\mathbf{\Sigma}
\ :=\ \langle \bm{n} \bm{n}^T \rangle \ = \ \left( \mathbf{B}^T \,
\mathbf{N}^{-1} \, \mathbf{B} \right)^{-1}.
\ee
Therefore, to obtain the ML map estimate one has to first compute the
inverse of the pixel to pixel noise covariance matrix
$\mathbf{\Sigma}^{-1} = \mathbf{B}^T \mathbf{N}^{-1} \mathbf{B}$.  The
ML map is then obtained as a solution to the linear algebraic equations, 
\begin{equation}
\mathbf{\Sigma}^{-1} \Pest =
\mathbf{B}^T \, \mathbf{N}^{-1} \cdot \mathbf{S}. 
\end{equation}

In solving this equation, we have a choice of either using one of a
number of direct methods or one of the iterative methods. In the low resolution
regime, the matrix inversion looks feasible with reasonable accuracy.
However, in general, direct
methods are computationally more expensive and iterative methods are
preferred, provided their convergence to the solution is rapid. For
sparse matrices the iterative Conjugate Gradient (CG) method is well
suited for this inversion~\cite{golub96,reid71} problem. The CG method
solves linear systems with symmetric positive definite matrix and have
been found to be more advantageous compared to other iterative methods
such as the Jacobi method~\cite{arm_wan04,cmbmap_comp}. Starting with
a guess solution, the convergence of the method can be often greatly
improved by `preconditioning' the system of equations. i.e.,
multiplying both sides with a suitable matrix (say, the inverse of
diagonal elements of the matrix.). Our choice is also motivated by the
fact that the CG method has been successfully implemented for map
making in CMB experiments~\cite{ned96,chal02, borrill99,jarosik06,cmbmap_comp}.

The above method can be extended to include multiple baselines and
also to estimate power in each polarisation component. We discuss
these extensions in the following subsections.

\subsection{Multiple Baselines}
\label{subsec:MB}

The above analysis can be extended to a set of $\Nb$ GW radiometer
baselines. Let $\mathbf{S}^{(i)}$ be the observed map by the $i^{th}$
baseline with beam matrix $\mathbf{B}^{(i)}$ and observed noise
$\mathbf{n}^{(i)}$. Then Eq.~(\ref{eq:conv}) can still be written as:
\begin{equation}
\mathbf{S} \ = \ \mathbf{B} \cdot \bm{\P} \ + \ \mathbf{n}, \nonumber
\end{equation}
where
\begin{eqnarray}
\mathbf{X} \ :=\ \left( \begin{array}{l}
\mathbf{X}^{(1)} \\ \mathbf{X}^{(2)} \\ \vdots \\ \mathbf{X}^{(\Nb)}
\end{array} \right)\,,\\\nonumber
\end{eqnarray}
with $\mathbf{X}$ representing the matrices $\mathbf{S}$,
$\mathbf{B}$, and $\mathbf{n}$. Note that $\mathbf{S}$, $\mathbf{n}$
are now $1\times \Npix\Nb$ vectors and $\mathbf{B}$ is a $\Npix\times
\Npix\Nb$ matrix, while $\bm{\P}$ (the true SGWB sky) remains
unchanged. This is similar to CMB experiments where each pixel is
visited by the detector several times. In the multi-baseline GW
radiometer case each pixel is visited by different baselines, and,
unlike a CMB experiment, each pixel is visited an equal number of
times. In this case too the Maximum Likelihood estimate\footnote{Of
course, one could combine the maps from different baselines with
suitable pixel dependent weight factors $\mathbf{w}_i$, $$\mathbf{S} =
\sum_i \mathbf{w}_i \cdot \mathbf{S}^{(i)},$$ which may also reduce the
noise by a factor of $\sim \sqrt{\Nb}$. But, it would have issues
of combining data from baselines with different beam
functions. Further analysis might be required to assess which method would
be more advantageous.} and the pixel to pixel noise covariance matrix
of the ML map are given by:
\be
\Pest \ = \ \mathbf{\Sigma} \,
\mathbf{B}^T \, \mathbf{N}^{-1} \cdot \mathbf{S}; \ \ \
\mathbf{\Sigma}^{-1} \ := \ \mathbf{B}^T \mathbf{N}^{-1}
\mathbf{B}, \label{eq:MLMap}
\ee
but the noise covariance matrix $\mathbf{N} := \langle \mathbf{n}\mathbf{n}^T \rangle$ of the raw sky map has to be modified. Let $n_i(\Om)$ be the pixel noise from the radiometer baseline $i$ with detectors $I$ and $I'$ and we follow the same convention for $\lambda_i$ and $Q_i$. Then,
\begin{widetext}
\begin{eqnarray}
\langle n_i(\Om_1) n_j(\Om_2) \rangle &=& \left[ \int \d t \, \lambda^{-1}_i(t,\Om_1) \right]^{-1} \left[\int \d t \, \lambda^{-1}_j(t,\Om_2) \right]^{-1} \int \d t_1 \, \int \d t_2 \, \nonumber\\
&& \int_{-\infty}^{\infty} \d f_1 \int_{-\infty}^{\infty} \d f_2 \int_{-\infty}^{\infty} \d f'_1 \int_{-\infty}^{\infty} \d f'_2 \, \delta_{\Delta t}(f_1-f'_1) \, \delta_{\Delta t}(f_2-f'_2) \ \times \nonumber\\
&& \langle \ft{n}_{I}^*(t_1,f_1) \ft{n}_{J}^*(t_2,f_2) \ft{n}_{I'}(t_1,f'_1) \ft{n}_{J'}(t_2,f'_2)\rangle \frac{\ft{Q}_i(\Om_1,t_1,f'_1;H)}{\lambda_i(t_1,\Om_1)} \frac{\ft{Q}_j(\Om_2,t_2,f'_2;H)}{\lambda_j(t_2,\Om_2)}.
\end{eqnarray}
\end{widetext}
If $i$ and $j$ denote the same baseline we get back the previous result  [Eq.~(\ref{eq:dirtyCov})]. However, if $i$ and $j$ denote different baselines, at least one of the two detector pairs will be different (i.e. either $I \ne J$ or $I' \ne J'$), so in that case $\langle n_i(\Om_1) n_j(\Om_2) \rangle = 0$. Hence, one can write
\begin{equation}
\langle n_i(\Om_1) n_j(\Om_2) \rangle \ = \ \delta_{ij} \, \langle n_i(\Om_1) n_j(\Om_2) \rangle \, .
\end{equation}
Thus, the matrix $\mathbf{N}$ for a network of baselines will be a block diagonal matrix, where each diagonal block $\mathbf{N}^{(i)}$ is the noise covariance matrix for the corresponding baseline, $N^{(i)}_{kk'} :=  \langle n_i(\Om_k) n_i(\Om_{k'}) \rangle$, provided in Eq.~(\ref{eq:dirtyCov}).

The above algebra suggests that it is fairly straight forward to combine the observations made by multiple baselines for the estimation of the true SGWB angular power distribution. In \cite{alnrmn, Malaspinas06} it was shown that the optimal way to search for a isotropic SGWB using a network of detectors (with uncorrelated noise) is to linearly combine correlations from pairs of detectors, instead of computing higher order correlations using data from more than two detectors. We can extrapolate the same logic to the directed search and argue that the procedure described above to combine data from a network of detectors is also optimal.

The search for a SGWB using a network of detectors is becoming progressively relevant as other kilometer scale detectors, namely Virgo, GEO and LCGT, are expected to reach their initial goal sensitivity in the next few years. A network of detectors can enhance the directed search in many ways. The resolution of a radiometer is proportional to the length of the baseline. Inclusion of a detector at a distance like Virgo, which is further away of the LIGO detectors than the mutual separation of the LIGO sites, will clearly increase the highest baseline separation and hence the resolution of the baseline. However, more important enhancement would be realized due to better coverage of the sky. An analogy with radio astronomy using an array of antennas may be appropriate to mention in this context \footnote{Notably, aperture synthesis technique using earth rotation was first introduced by Martin Ryle for radio observation of cosmic sources.}. As the earth rotates, the projections of the radio antenna baselines on the plane perpendicular to the source direction sample different points on the two dimensional Fourier plane (commonly known as the $u$-$v$ plane). The sampled Fourier plane is then inverse transformed to generate the image. While the {\em highest resolution} of the network is limited by the projection of maximum antenna separation, addition of more antennas to the network ensures better sampling of the $u$-$v$ plane reducing the side lobes,  thereby producing a more faithful image of the sky. Detailed introduction to the basic principles of earth rotation aperture synthesis can be found in most of the standard texts on radio astronomy, e.g., \cite{radioastron}. In GW radiometry with a network of detectors we expect that a similar scenario will arise - better coverage of the sky should be possible due to different orientations of the baselines with respect to the source. Moreover, since the true power distribution will be estimated from an over-constrained set of equations, the error in the estimated quantities will be reduced. In addition, a radiometer search can benefit from certain technical advantages that a network of detectors can provide: A detector at a third site joining the LIGO detectors will boost the ``single-baseline integration time'', i.e., the single-baseline duty-cycle in a three-site network will be at least as good as, but will likely be better than, that in a two-site network. Also, owing to common instrumental noise sources in the LIGO detectors, certain frequency bands are currently notch-filtered in computing the cross-correlation statistic. Some of these noise sources are known not to affect Virgo and, therefore, the LIGO-Virgo cross-correlation statistics. For example, the power-line noise affects the LIGO detectors at the harmonics of 60Hz, whereas it affects the Virgo detector at the harmonics of 50Hz. Therefore, a radiometer search that benefits from the LIGO-Virgo baseline's contribution will probe the presence of astrophysical signals over a larger set of frequencies than one limited to the baseline consisting of the LIGO pair of detectors.

\subsection{Polarization Map}

We may also choose to extract power from different polarizations separately. The discrete convolution equation,
\begin{equation}
\mathbf{S} \ = \ \mathbf{B}_+ \cdot \bm{\P}_+ \ + \ \mathbf{B}_\times \cdot \bm{\P}_\times \ + \ \mathbf{n} \,,
\end{equation}
can also be expressed by Eq.~(\ref{eq:conv})
\begin{equation}
\mathbf{S} \ = \ \mathbf{B} \cdot \bm{\P} \ + \ \mathbf{n}, \nonumber
\end{equation}
where
\begin{equation}
\mathbf{B} \ := \ \left[ \begin{array}{ll}
\mathbf{B}_+ & \mathbf{B}_\times
\end{array} \right]; \ \ \
\bm{\P} \ := \ \left( \begin{array}{l}
\bm{\P}_+ \\ \bm{\P}_\times
\end{array} \right).\label{eq:defPolMap}
\end{equation}
Again in this case the ML estimator of sky map can be expressed by Eq.~(\ref{eq:MLMap}):
\be
\Pest \ = \ \mathbf{\Sigma} \,
\mathbf{B}^T \, \mathbf{N}^{-1} \cdot \mathbf{S}; \ \ \
\mathbf{\Sigma}^{-1} \ := \ \mathbf{B}^T \mathbf{N}^{-1}
\mathbf{B}. \no
\ee

This case can also be generalized for a network of detectors by retaining the same definition of $\bm{\P}$ [Eq.~(\ref{eq:defPolMap})], but redefining the beam matrix as:
\begin{equation}
\mathbf{B} \ := \ \left[ \begin{array}{ll}
\mathbf{B}_+^{(1)} & \mathbf{B}_\times^{(1)}\\ \mathbf{B}_+^{(2)} & \mathbf{B}_\times^{(2)}\\ \vdots & \vdots \\ \mathbf{B}_+^{(\Nb)} & \mathbf{B}_\times^{(\Nb)} 
\end{array} \right] \,
\end{equation}
and, again, using the same ML estimation formula given in Eq.~(\ref{eq:MLMap}).

\section{Implementation and Numerical Results}
\label{sec:numRes}

In this work we have numerically implemented the maximum likelihood estimation algorithm on simulated data using the MATLAB\textregistered~ software package~\cite{MATLAB} to estimate the ``true'' (unpolarized) SGWB sky observed with a single baseline ground based GW radiometer. The details of the computation scheme, the numerical deconvolution algorithm, the simulated data and the deconvolved maps are presented in this section.

\subsection{Preparation of Simulated Dirty Maps}

The data are simulated in the frequency domain for each time segment. Since the noise of ground based interferometric detectors is very high at frequencies greater than few $100$~Hz and the computation cost increases with the number of frequency bins, we use an upper cut-off frequency of $f_u = 512$~Hz and bin width of $\Delta f = 2$~Hz for testing of the algorithm. The length of each time segment is chosen as $\T = 192$~sec and the total integration time is $T = 86400$~sec\footnote{For convenience we have used $86400$~sec for one sidereal day, instead of $86164$~sec. It hardly affects the accuracy of the results presented in this paper, as the same duration has been used for both injecting signals and analyzing data. Even for analyzing real data {\em the same segments can be used}, however, earth's rotation frequency should be accurately supplied in order to establish correct correspondence between time and celestial coordinates.}. The sky is pixelized using the Hierarchical Equal Area isoLatitude Pixelization (HEALPix)\footnote{\url{http://healpix.jpl.nasa.gov/}} scheme, which divides the $2$-sphere ($S^2$) in $12 \, n_\text{side}^2$ pixels, where $n_\text{side}$ is an integer power of $2$. Since the radiometer beam-width is greater than $\sim 6^\circ$, we chose $n_\text{side} = 16$, which corresponds to a pixel width of $\sim 3^\circ$ and a total of $3072$ pixels. The HEALPix scheme also allows fast spherical harmonic transform on a sphere, which may become useful for more advanced analysis in future. Note that, the algorithm is {\em independent} of the pixelization scheme, other equal area pixelization schemes can also be used in the analysis.

We generate the detector noise $\ft{n}_I(t;f)$ using a Gaussian pseudo random number generator for each time segment. The noise is colored using the (one sided) noise PSD $P_I(t;f)$ of the corresponding detector according to Eq.~(\ref{eq:GWChunkNoise}). MATLAB\textregistered~ software's pseudo random number generator {\tt randn} can generate very long sequences of random numbers, so we relied on that routine for simulating detector noise. For each of $T/\T = 86400/192 = 450$ time  segments, we generated a complex random sequence (that is, two real random sequences) of $f_u/\Delta f = 512/2 = 256$ real numbers. The total number of random numbers, $2 (T/\T)(f_u/\Delta f) = 225,000$, is much less than the period of {\tt randn}, which is $2^{1492} \gtrsim 10^{449}$~\cite{randMATLAB}.

Signal is also generated directly in the frequency domain. However, the GW strain in each detector, $\ft{h}_I(t;f)$, is not generated independently; rather the product of the strains in the detectors, $\ft{h}_1^*(t;f) \, \ft{h}_2(t;f)$, is generated directly using the statistical properties of the strain correlation described in subsection~\ref{TimeFreqData}, in particular, Eq.~(\ref{h1h2}).
We may write the product of the strains as a sum of its expectation value and statistical fluctuations:
\begin{equation}
\ft{h}^*_1(t,f) \, \ft{h}_2(t,f) \ = \ \langle \ft{h}^*_1(t,f) \, \ft{h}_2(t,f) \rangle \ + \ \text{fluctuations}.
\label{eq:sigFluc}
\end{equation}
Since our main aim is to generate,
\begin{equation}
\ft{s}^*_1(t,f) \, \ft{s}_2(t,f) = [\ft{h}^*_1(t,f) + \ft{n}^*_1(t,f)] [\ft{h}_2(t,f) + \ft{n}_2(t,f)]
\end{equation}
and since statistically the variation in the signal terms are much weaker than the {\em zero mean uncorrelated} detector noise terms, we may simply drop the signal ``fluctuations'' term from Eq.~(\ref{eq:sigFluc}) - that is, we may approximate the product of the detector outputs using the formula\footnote{Note that, it is also possible to generate the correlated detector strains $h_I(t)$ independently and construct $\ft{s}_I(t;f) = \ft{h}_I(t;f) \, + \, \ft{n}_I(t;f)$ for each detector $I$ separately using simulated colored noise $\ft{n}_I(t;f)$ to test the analysis. However, we chose the above method in order to reduce complications in the primary testing of the analysis presented in this paper.
}:
\begin{equation}
\ft{s}^*_1(t,f) \, \ft{s}_2(t,f) \ = \ \langle \ft{h}^*_1(t,f) \, \ft{h}_2(t,f) \rangle \ + \ \ft{n}^*_1(t,f) \, \ft{n}_2(t,f). \label{eq:simulateGWBStrainCorr}
\end{equation}
For all the cases considered in this paper we have used flat source PSDs, i.e., $H(f) = \text{constant}$.

In this analysis we assume the sky to be a collection of {\em uncorrelated point sources} of different strengths placed at every pixel. Moreover, the numerical analysis has been restricted to the case of equal power in each polarization. So the (injected) true sky is constructed by putting
\begin{equation}
\P_\text{true}(\Om) \ = \ \sum_k \P_k \, \delta(\Om - \Om_k),
\end{equation}
where $\P_k$ is the strength of the point source placed at pixel $k$, located in the direction $\Om_k$ (in order to inject only one point source at pixel $k_0$, we make all the $\P_k = 0$ except for $k=k_0$). In this set up, the expression for the overlap reduction function [Eq.~(\ref{eq:gamma})] for the true SGWB strain becomes
\begin{equation}
\gamma(t, f; \dom, d\P_A) \ = \ \sum_{k} \Gamma(\Om_k,t) \, \P_k \, e^{2\pi \i f \Om_k\cdot\Dx(t)/c},
\end{equation}
where we use our usual notation for $\Gamma(\Om,t)$ defined by Eq.~(\ref{Gmm}).

We substitute the above in Eq.~(\ref{h1h2}) and inject that simulated signal in noise using Eq.~(\ref{eq:simulateGWBStrainCorr}) to generate products of outputs from two detectors. In order to preserve the reality of time series data, the products of signals are generated only for positive frequencies and setting the negative frequencies equal to the complex conjugates of their positive frequency counterparts, that is, $\ft{s}_I(t,-f) = \ft{s}^*_I(t,f)$.

The radiometer analysis is then run on the simulated data by placing filters $\ft{Q} (\Om_k, t, f; H)$ at each pixel $k$ to generate the dirty maps.

\subsection{Deconvolution: Clean Maps}

Any deconvolution routine requires the beam function at all the points
of interest on the sky. For a GW radiometer, the filters (and hence
the beam functions) are dependent on the data set itself. So, if the beam
function is calculated for each sky pixel, apparently the computational 
cost should go up by a factor of number of pixels ($\Npix \sim 3000$)
times the cost to make one sky map of beam for a given pointing direction.
However, we can use
algebraic tricks to make this method computationally viable. A
possible way to implement this method for the simple case of one
baseline and equal power in each polarization is demonstrated below.

The beam and noise covariance matrices are given by:
\begin{widetext}
\begin{eqnarray}
\label{eq:Bij}
B_{ij} &=& 2 \Delta f \left[ \sum_{t=0}^T \frac{1}{\lambda(t,\Om_i)} \right]^{-1} \sum_{t=0}^T \Gamma(\Om_j,t) \, \Gamma(\Om_i,t) \, \Re \left[ \sum_{f=0}^{f_u}  e^{2\pi i f (\Om_j-\Om_i)\cdot\Dx(t)/c} G(t,f) \right]\,,\\
N_{ij} &=& \frac{\Delta t}{4}\left[ \sum_{t=0}^T \frac{1}{\lambda(t,\Om_j)} \right]^{-1} B_{ij} \,, 
\end{eqnarray}
\end{widetext}
where
\bea
\Gamma(\Om,t) &:=& F^+_1(\Om,t) F^+_2(\Om,t) + F^\times_1(\Om,t) F^\times_2(\Om,t)\,,\no\\
G(t,f) &:=& H^2(f)/[P_1(t,f) P_2(t,f)] \,.
\eea
Now one can see that the beam matrix is a summation of parts which
depend on either $\Om_i$ or $\Om_j$. So, it is possible to precompute
the arrays $\lambda(\Om,t)$, $\Gamma(\Om,t)$, $G(t,f)$ and $\Dx(t)$, only 
once, and then use them efficiently to evaluate the whole $\mathbf{B}$ matrix.
Since each of these arrays are functions of {\em any two} of the three variables
$\Om$,$t$,$f$, the memory size required for each two dimensional array will not be too
large. Once the arrays are computed, the next step is to obtain each element of
the beam matrix separately. Each element requires the evaluation of two sum loops (like matrix multiplication) involving one exponential. The symmetry of the beam matrix (without the
normalization constant) can be utilized here to reduce the computation cost
by a factor close to $2$.
To summarize, a significant amount of CPU time required
to make a skymap of the beam for one pointing direction could be utilized to make
the beam maps for all other pointing directions. This is true for the noise
covariance matrix as well, which, in this case, is proportional to
the (unnormalized) beam matrix. A Fast Fourier Transform (FFT) with interpolation trick~\cite{ballmer06} and
assumption of stationarity of noise for deconvolution (\emph{not} for
making the dirty map, where non-stationarity will be
accounted for) can also be incorporated in future to reduce the CPU time.

In this simple case, since the beam matrix is a square matrix, so that, $\mathbf{\Sigma} \, = \, \left( \mathbf{B}^T\mathbf{N}^{-1} \mathbf{B} \right)^{-1}  \, = \, \mathbf{B}^{-1} \mathbf{N} (\mathbf{B}^{-1})^T$, the estimated map given by Eq.~(\ref{eq:MLMap}) is just the least square solution:
\begin{equation}
\Pest \ =\ \mathbf{B}^{-1} \cdot \mathbf{S}. \label{eq:leastSq}
\end{equation}
Even for the general cases of multiple baselines and polarized background, the estimation equation takes the above simple form [see section~\ref{sec:ML}].

\begin{figure}[h]
\centering
\includegraphics[width=0.5\textwidth]{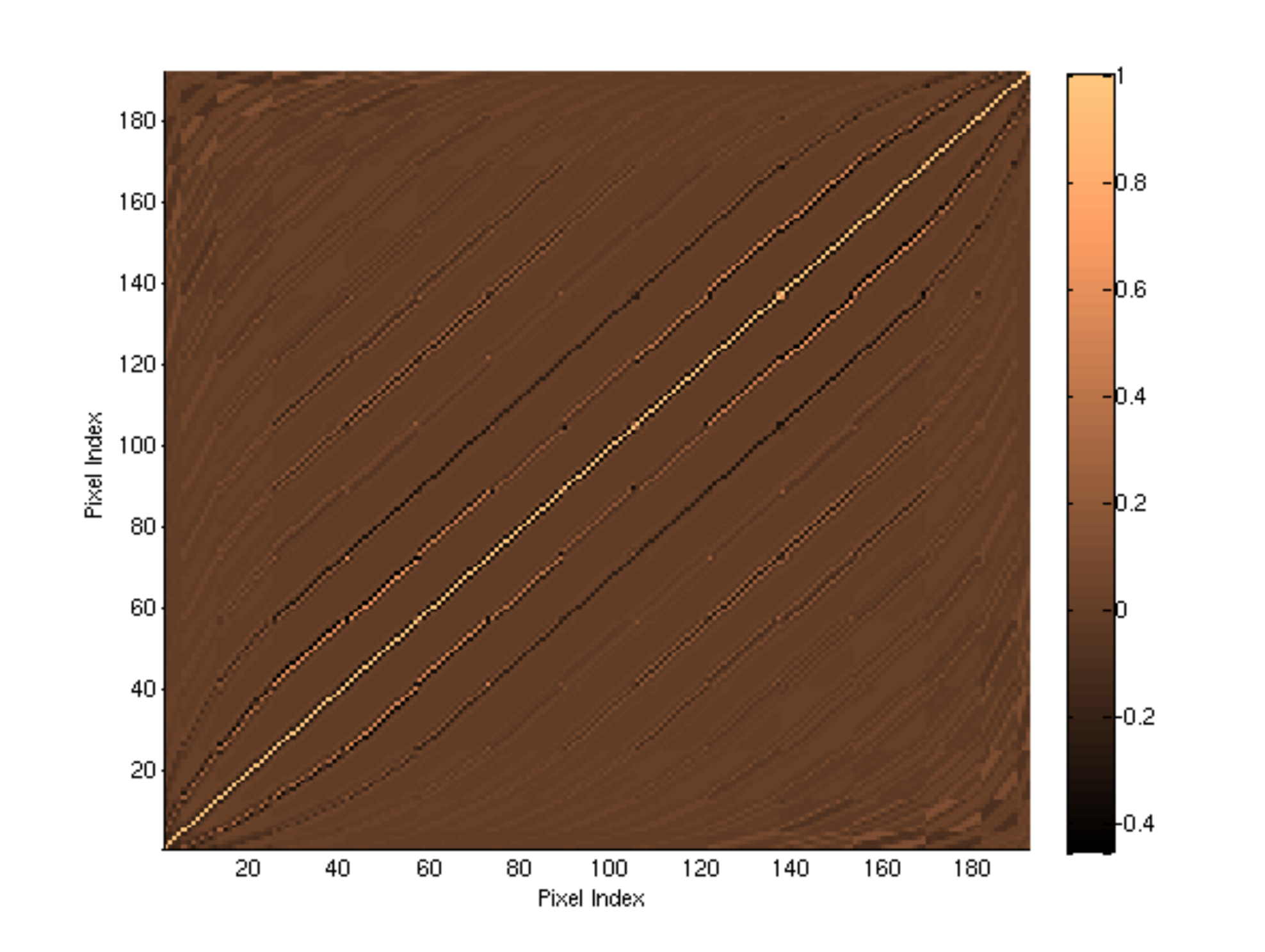}
\caption{A typical Beam matrix for the LIGO Hanford-Livingston baseline at a low resolution ($192$ pixels) is shown in this figure. Each row of the matrix is the beam response function for the pointing direction that corresponds to the row index. The matrix is diagonal dominated as the radiometer receives maximum contribution from the pointing direction. The stripes are related to the isoLatitude pixelization scheme - the indices of the neighboring pixels at different latitudes differ by the total number of pixels on that latitude. The possibility of making the beam matrix more diagonal dominated using a nested pixelization scheme, where the indices of the neighboring pixels are close, is being explored.}
\label{fig:GWBBeamMat}
\end{figure}

The first task was to compute the beam matrix, which happens to be the computationally most intensive task. A typical beam matrix for the LIGO baseline using $192$ HEALPix pixels is shown in Figure~\ref{fig:GWBBeamMat}. By construction $B_{kk} = 1$ and $|B_{kk'}| < 1$ for $k \ne k'$, hence the matrix is diagonal dominated. The ``stripes'' in the matrix are related to the pixelization scheme. The beam is stronger if the pixels are closer to the pointing direction and it weakens as the distance between the pixels and pointing direction increases. In other words, the pixels closer to a point source will have stronger contamination from the point source. However, since we have used a isoLatitude pixelization scheme, the indices of two neighboring pixels at different latitudes differ by the total number of pixels on that latitude. This fact is reflected in the plot of the beam matrix - the matrix is sparse with certain ``periodic'' behavior which  produces the stripes in the plot. The matrix becomes even more sparse for finer resolutions as greater number of isoLatitude pixel rings pass through the core of the beam. Making a legible plot of the beam matrix for higher resolution is difficult, so the plot presented here is restricted to lower resolution - $192$ pixels instead of $3072$.

Since the sparseness of the beam matrix depends on the pixelization scheme, it may be possible to make the beam matrix significantly diagonal by using a nested pixelization scheme, where the indices of the neighboring pixels are close. This possibility is being explored.

Sparse matrices are computationally easier to invert; however, the stability of inversion of such matrices is a numerical challenge. Therefore, as mentioned in section~\ref{sec:ML}, instead of evaluating $\Pest=\mathbf{B}^{-1} \cdot \mathbf{S}$ [Eq.~(\ref{eq:leastSq})], we choose to algebraically solve for $\Pest$ from the set of linear equations
\begin{equation}
\mathbf{B} \cdot \Pest \ = \ \mathbf{S},
\end{equation}
with the same number of unknowns as the number of equations (i.e.,  the system is {\em not} under or over constrained). The above equation can be cast into a linear system with symmetric kernel,
\begin{equation}
\bm{\mathfrak{b}} \cdot \Pest \ = \ \bm{\mathfrak{s}}, \label{eq:numEstEqn}
\end{equation}
by introducing two new quantities,
\begin{equation}
\mathfrak{b}_{ij} \ := \ \left[ \sum_{t=0}^T \frac{1}{\lambda(t,\Om_i)} \right] \, B_{ij}; \ \mathfrak{s}_i \ := \ \left[ \sum_{t=0}^T \frac{1}{\lambda(t,\Om_i)} \right] \, S_i.
\end{equation}
Comparing with Eq.~(\ref{eq:Bij}) one can clearly see that $\bm{\mathfrak{b}}$ is a symmetric matrix. A symmetric kernel is always preferred by most of the algorithms for solving numerical linear equations. Moreover, it is possible to show from the algebra presented in subsection~\ref{subsec:MB} that, to incorporate observations from multiple baselines, one can simply replace $\bm{\mathfrak{s}}$ and $\bm{\mathfrak{b}}$ by the sum of those respective quantities over all the baselines and solve Eq.~(\ref{eq:numEstEqn}) for $\Pest$ (which is, of course, independent of any baseline). Thus the extension of this analysis to incorporate a network of detectors becomes straight forward with these new quantities.

Following the discussion presented in section~\ref{sec:ML}, we use a Conjugate Gradient (CG) iterative technique to solve the above set of linear equations. CG algorithm solves a set of linear equations  $\mathbf{A} \cdot \mathbf{x} \ = \ \mathbf{b}$, where $\mathbf{A}$ is a square matrix and $ \mathbf{x}$, $ \mathbf{b}$ are vectors, by minimizing the quadratic form ${1 \over 2} \, \mathbf{x} \cdot \mathbf{A} \cdot \mathbf{x} \ - \ \mathbf{b} \cdot \mathbf{x}$. We use the {\em minimum residual} method, which efficiently utilizes the fact that $\bm{\mathfrak{b}}$ is symmetric and does not require $\bm{\mathfrak{b}}$ to be positive definite. The minimum residual method aims to minimize the residual $|\mathbf{A} \cdot \mathbf{x} - \mathbf{b}|^2$ itself, instead of the quadratic form ${ 1 \over 2} \mathbf{x} \cdot \mathbf{A} \cdot \mathbf{x} - \mathbf{b} \cdot \mathbf{x}$. Further details on CG methods can be found in standard literature, e. g., \cite{numrec94}.

The clean maps also contain pixel noise - partly due to the random noise present in the data and partly due to the numerical errors introduced at each stage of the pipeline, mainly during the process of deconvolution. There are pixels in the deconvolved map, which have negative values, even though the injected map is positive. To reduce the noise in the clean maps, we introduce an additional step: We compute the root-mean-square (RMS) noise ``$\sigma$'' in a map when there is no injected source. Then in the clean map (with source) we {\em mask}, that is, set to zero, all the pixels that have values less than a threshold of few $\sigma$. The number of iterations for deconvolution and the threshold for masking can be adjusted according to the tolerable levels of false alarm and false dismissal probabilities.

The above can be easily extended to handle real data where we have no control on the injections. One can calculate the equivalent of RMS noise for no injection by shifting the data streams from different detectors by a large time lag (much smaller than the segment duration), say, $1$~sec, that corresponds to distances much greater than the earthly distances, so that, true GW signals are not added coherently. To be more careful, one can perform this exercise for a few large time shifts and confirm that the noise levels are not significantly different for different shifts.

It is, however, not so straight forward to measure the quality of deconvolution. The signal-to-noise ratio (SNR) of individual pixels do not carry enough significance, as the neighboring pixels are highly correlated. It is also difficult to define a quantity that can take into account the pixel-to-pixel noise covariance due to the difficulty in inverting the beam matrix. In this paper, we use a rather simplistic measure to quantify the quality of deconvolution, which is often used in image processing to measure the reconstruction error. We use a quantity known as the ``Normalized Mean Square Error" (NMSE)~\cite{NMSE}, expressed in terms of the injected $\bm{\P}$ and the estimated $\Pest$ maps as
\begin{equation}
\text{NMSE} \ := \ \frac{| \Pest - \bm{\P} |^2}{|\bm{\P}|^2}.
\end{equation}
Obviously, lower the NMSE, better the reconstruction.

The whole analysis was tested for different kinds of injected maps consisting of localized sources and diffuse sources. In all these cases each pixel $k$ of the injected map was assigned a value $\mathcal{P}_k$ between 0 and 1 with a source PSD $H(f) = 5\times 10^{-47}/\text{Hz}$. This means that, if a pixel of a test map has strength $1$, the standard deviation of the Fourier transform of stochastic GW coming from that pixel is $\sqrt{H(f)} \sim 7 \times 10^{-24}/\sqrt{Hz}$. This standard deviation is about one third of the standard deviation of Fourier transform of noise at the most sensitive frequency band of the LIGO-I detectors which is about $2-3\times10^{-23}/\sqrt{Hz}$. To our knowledge, the strength of anisotropic astrophysical GW background has not so far been predicted theoretically. However, if we try to extend the results from the all-sky averaged (isotropic) astrophysical background~\cite{CowardTania06} to have a crude estimate of the strength of the anisotropic background, it turns out that, the PSD of the anisotropic astrophysical background in the universe is weaker than the $H(f)$ we have injected by roughly a few orders of magnitude. In the present work we have used an observation time of {\em one day} to demonstrate the method. With longer observation times and employing several baselines comprising of the upcoming advanced (more sensitive) detectors, the difference between the expected background and the detectable background would diminish or altogether disappear.

\begin{figure}
\subfigure[~Injected map]{\label{fig:PtMap-a}\includegraphics[width=0.415\textwidth]{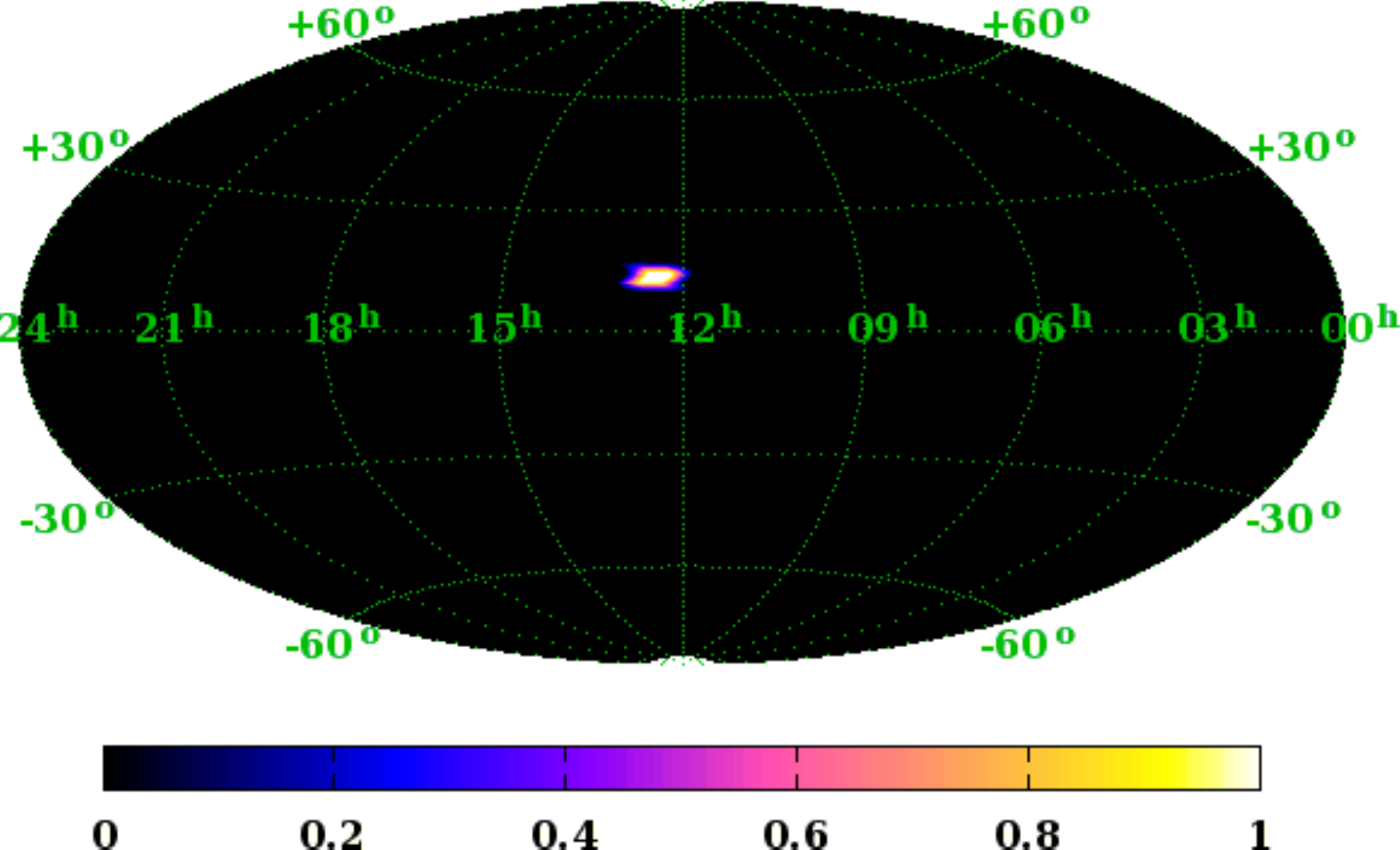}}\\
\subfigure[~Dirty map - made from simulated data using the radiometer analysis]{\label{fig:PtMap-b}\includegraphics[width=0.415\textwidth]{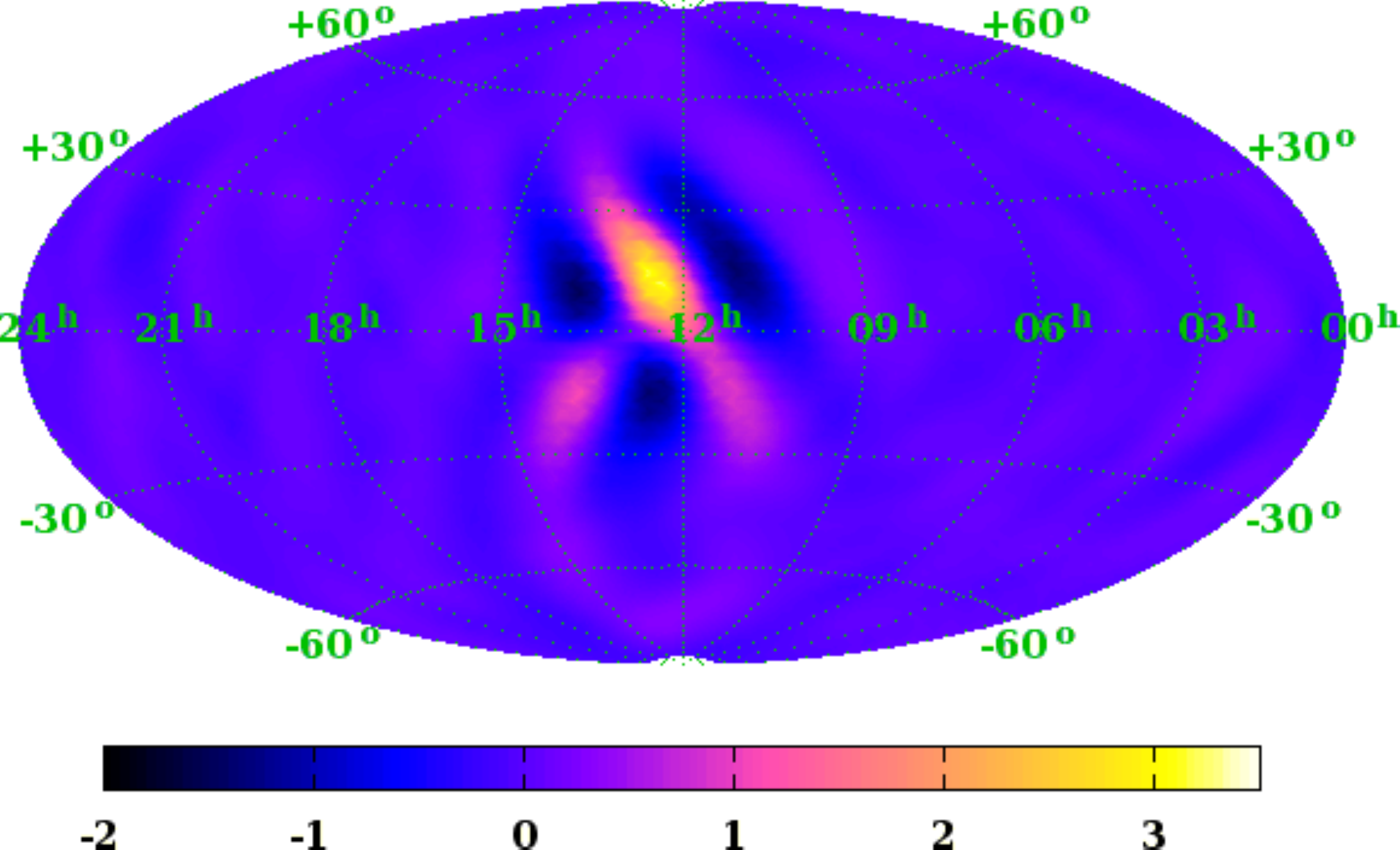}}\\
\subfigure[~Clean map - obtained by deconvolution of the dirty map using $15$ CG iterations (NMSE = $1.22$)]{\label{fig:PtMap-c}\includegraphics[width=0.415\textwidth]{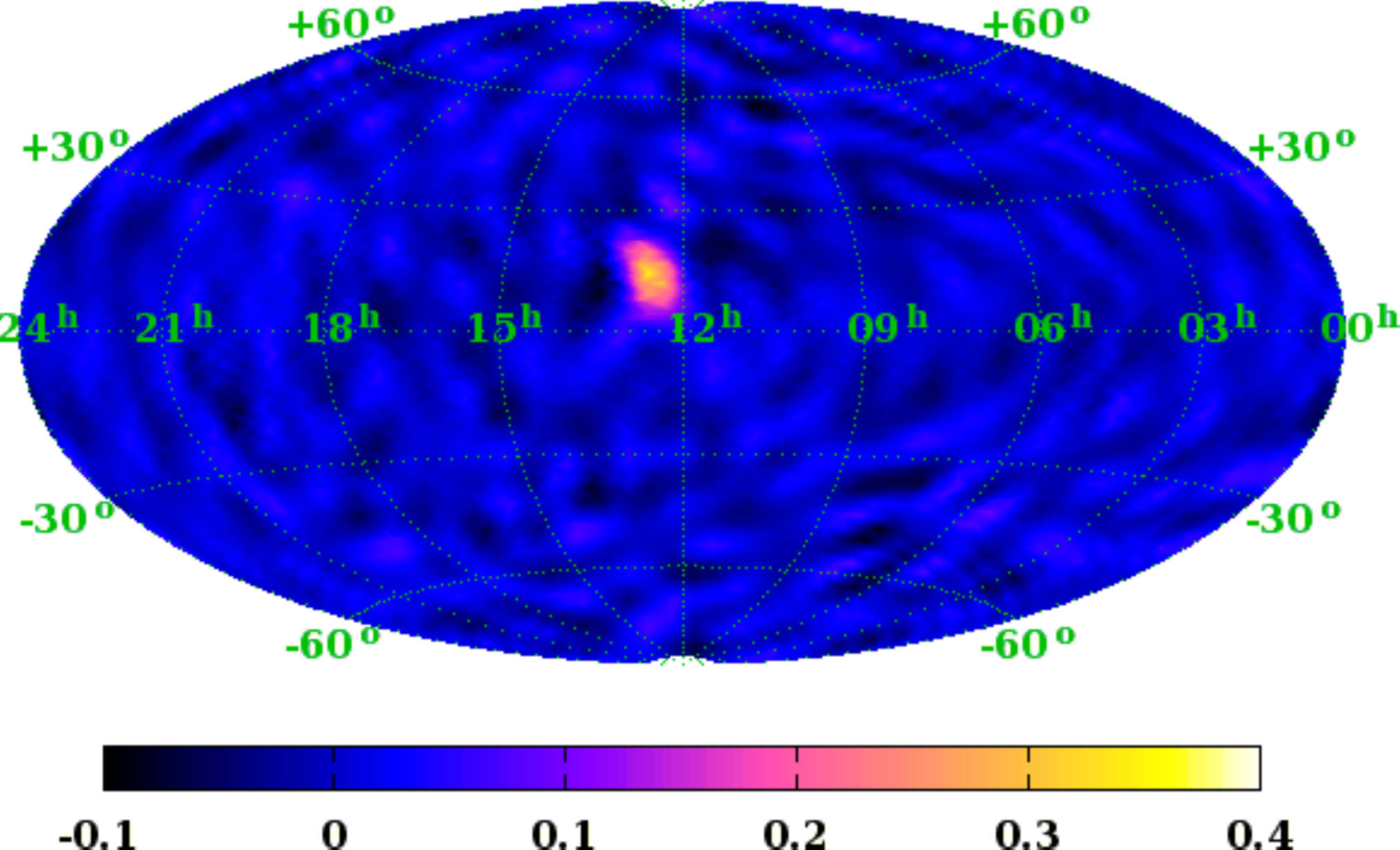}}\\
\subfigure[~Masked clean map - all the pixels below a threshold of $5\sigma$ were set to zero (NMSE = $0.64$)]{\label{fig:PtMap-d}\includegraphics[width=0.415\textwidth]{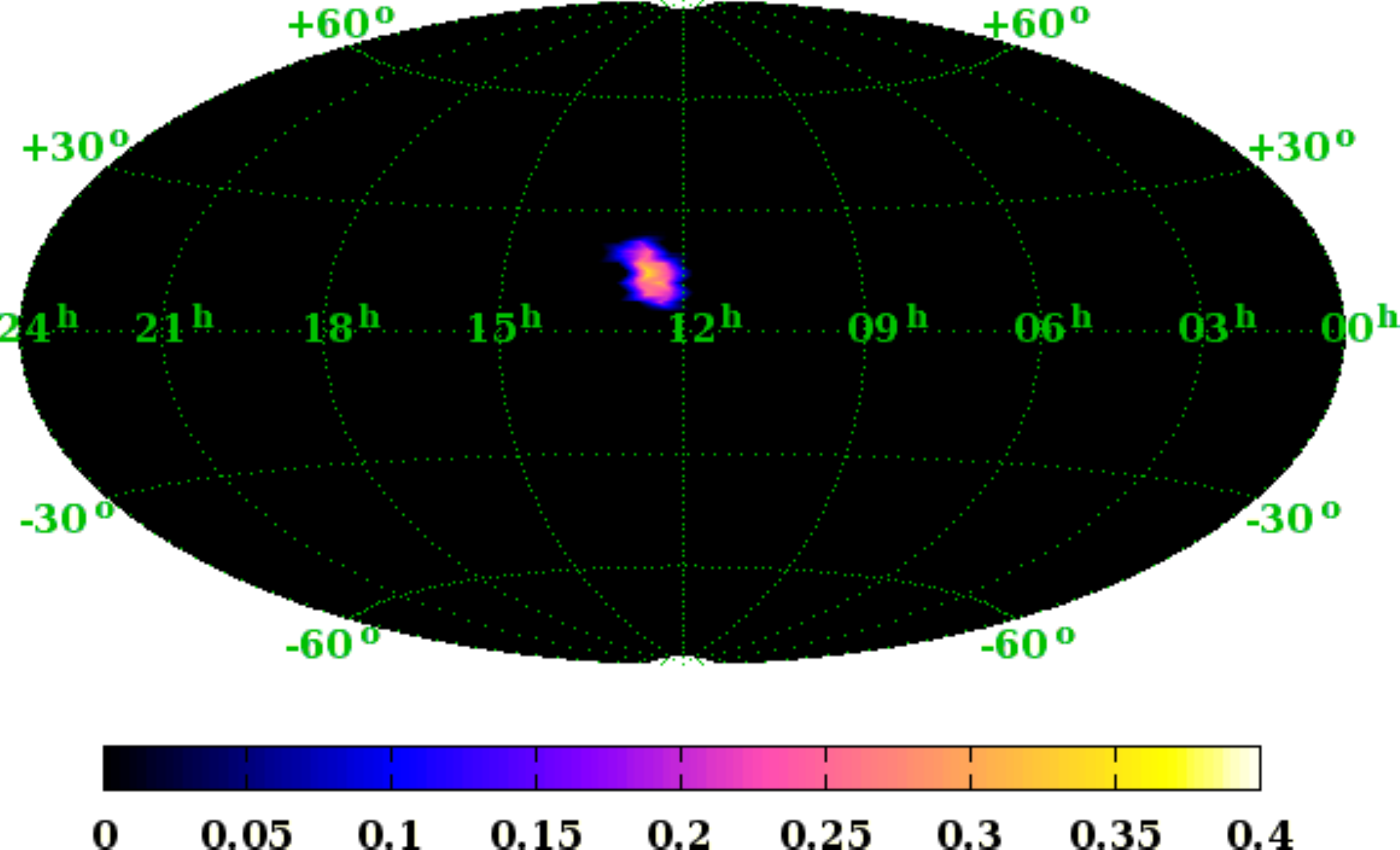}}
\caption{Illustration of deconvolution for localized sources: A $4$-pixel wide localized source was injected near the location of the Virgo cluster - a potential source of SGWB.}
\label{fig:PtMap}
\end{figure}

We first inject a $4$-pixel wide localized source near the Virgo cluster, a potential point source of SGWB, as illustrated in Figure~\ref{fig:PtMap}. Figure~\ref{fig:PtMap-a} shows the injected map. Figure~\ref{fig:PtMap-b} shows the dirty map made from simulated data. Figure~\ref{fig:PtMap-c} shows the clean map, obtained by deconvolving the above dirty map with the beam using $15$ conjugate gradient iterations and Figure~\ref{fig:PtMap-d} shows the same clean map masked using a $5\sigma$ threshold. It is evident that deconvolution has successfully localized the source in a relatively smaller area as compared to the dirty map. Still, one should note that, the deconvolution routines do not perform well when the injected source is like a delta function. This causes high NMSE, $1.22$ for the unmasked and $0.64$ for the masked clean maps, and significant loss of the peak strength of the reconstructed point source, as indicated in Figure~\ref{fig:PtMap}. Moreover, increasing the number of iterations beyond a certain level actually deteriorates the quality of deconvolution due to noise amplification, and this level is dependent on the kind of source one is searching for. In this basic analysis we have used $15$ iterations to search for localized sources and $40$ iterations to search for broad sources, which offer reasonably clean deconvolution and comparatively low NMSE. Introduction of a minimum error criterion~\cite{Tang} to terminate the iteration process is being considered. Several other deconvolution algorithms are being explored in order to identify the one which is {\em best} suited for the GW radiometer analysis.

\begin{figure*}
\centering
\subfigure[~Injected maps]{\label{fig:DiffMap-a}%
\includegraphics[width=0.415\textwidth]{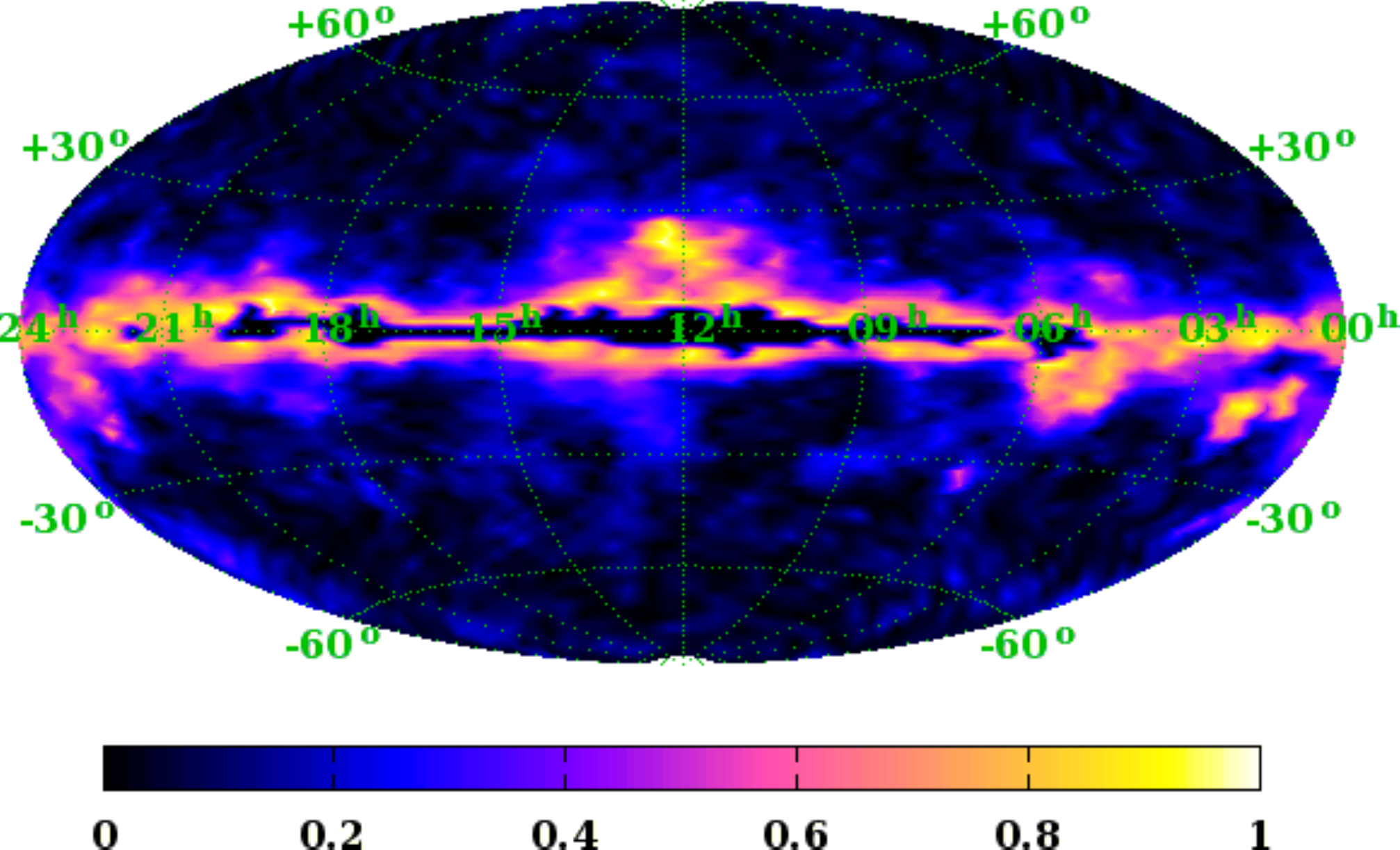}\qquad\qquad%
\includegraphics[width=0.415\textwidth]{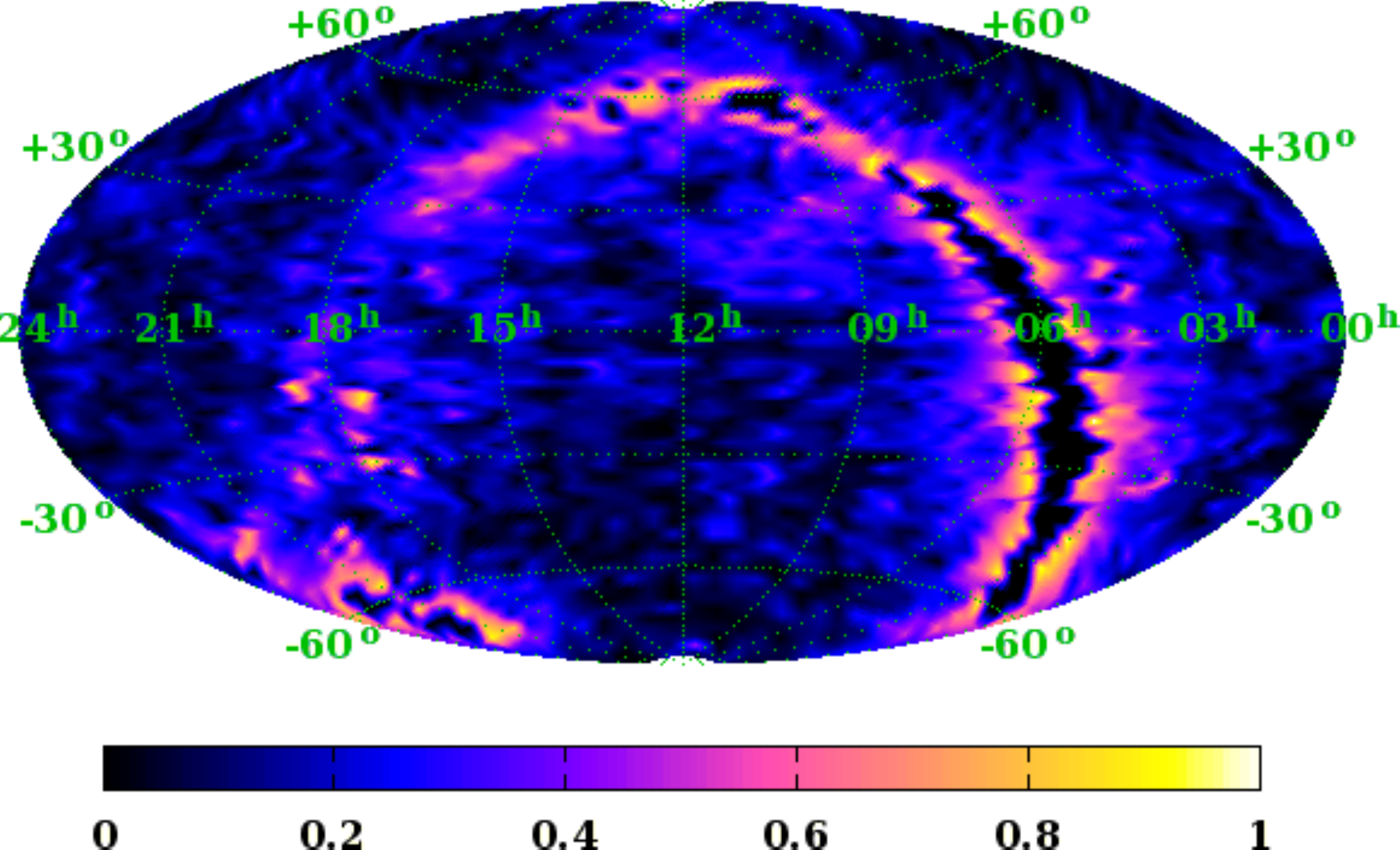}}\\
\subfigure[~Dirty maps - obtained from simulated data using the radiometer analysis]{\label{fig:DiffMap-b}%
\includegraphics[width=0.415\textwidth]{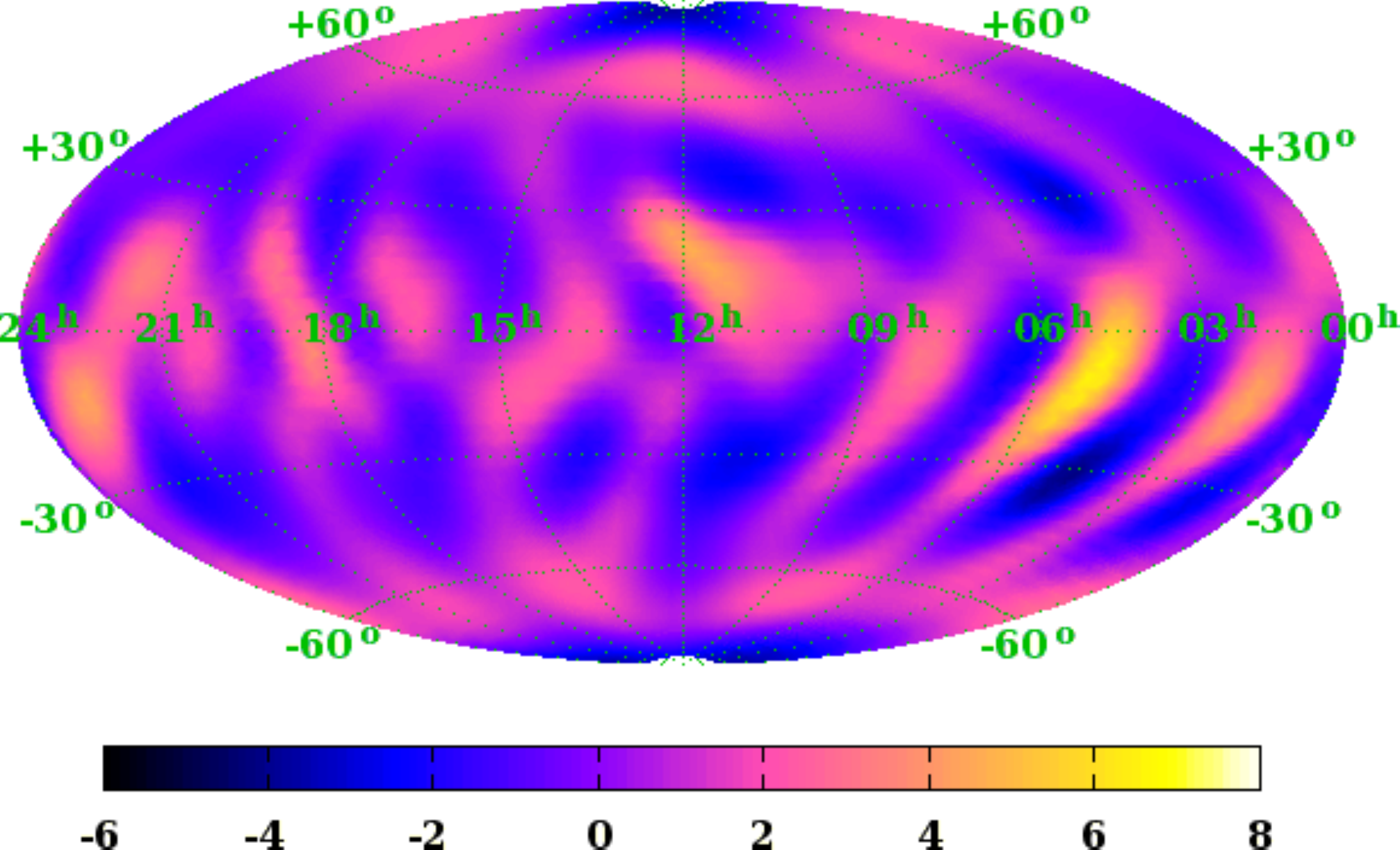}\qquad\qquad%
\includegraphics[width=0.415\textwidth]{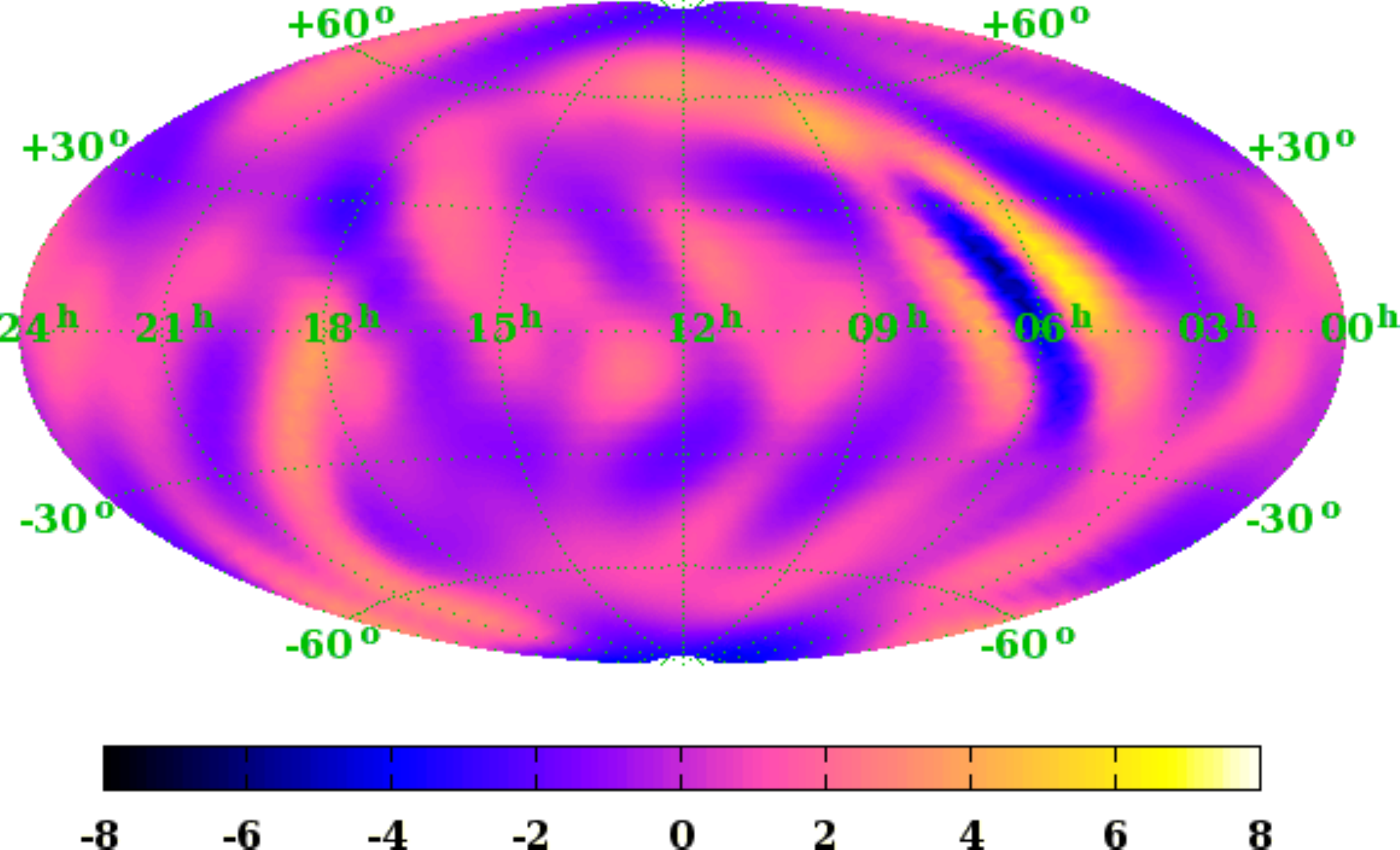}}\\
\subfigure[~Clean maps - obtained by deconvolution of the dirty maps using $40$ CG iterations. Clearly, the structures and positivity of the injected maps, which were lost in the dirty maps, have been restored quite significantly.\newline(left panel: NMSE = $0.33$; right panel: NMSE = $0.22$)]{\label{fig:DiffMap-c}%
\includegraphics[width=0.415\textwidth]{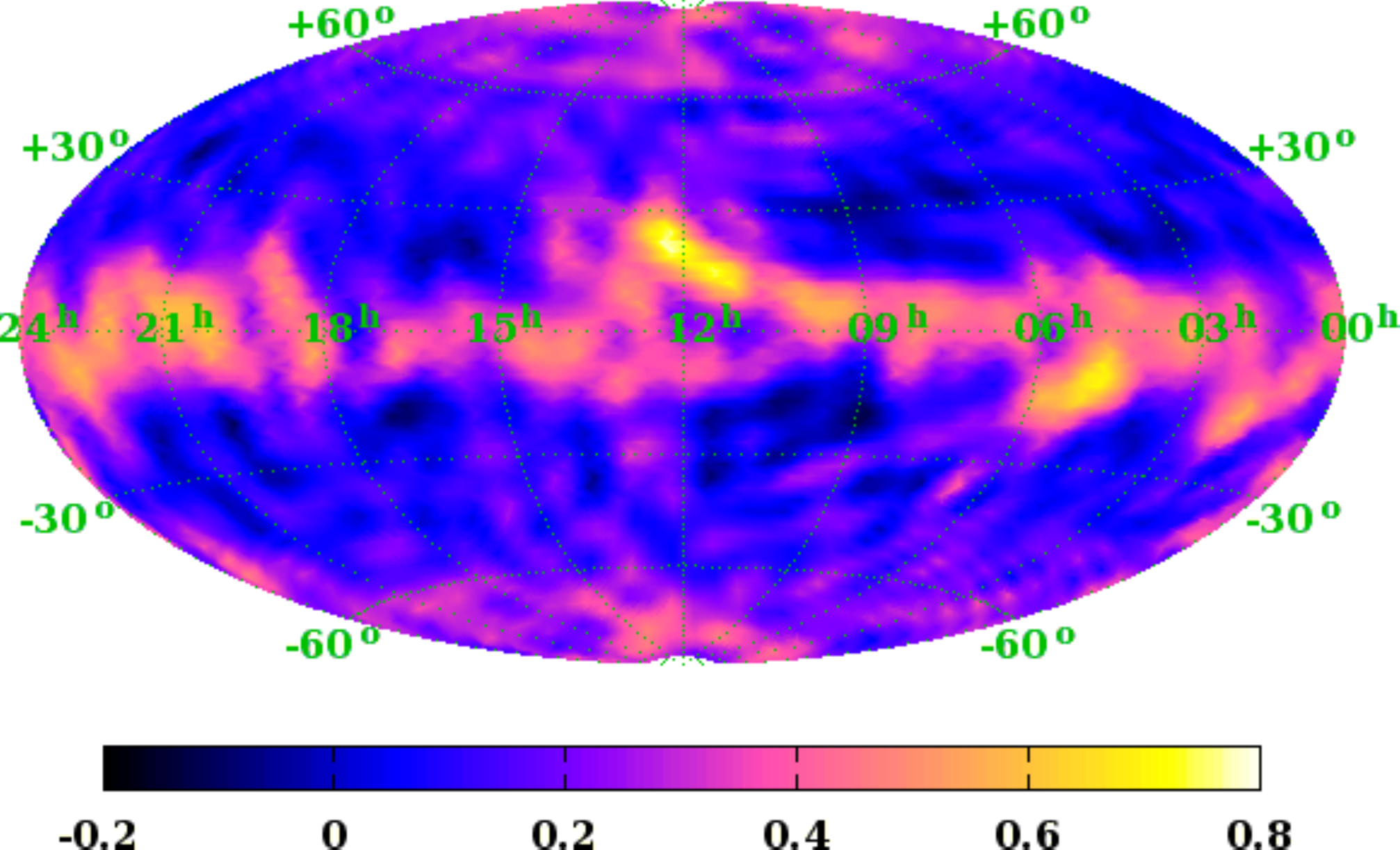}\qquad\qquad%
\includegraphics[width=0.415\textwidth]{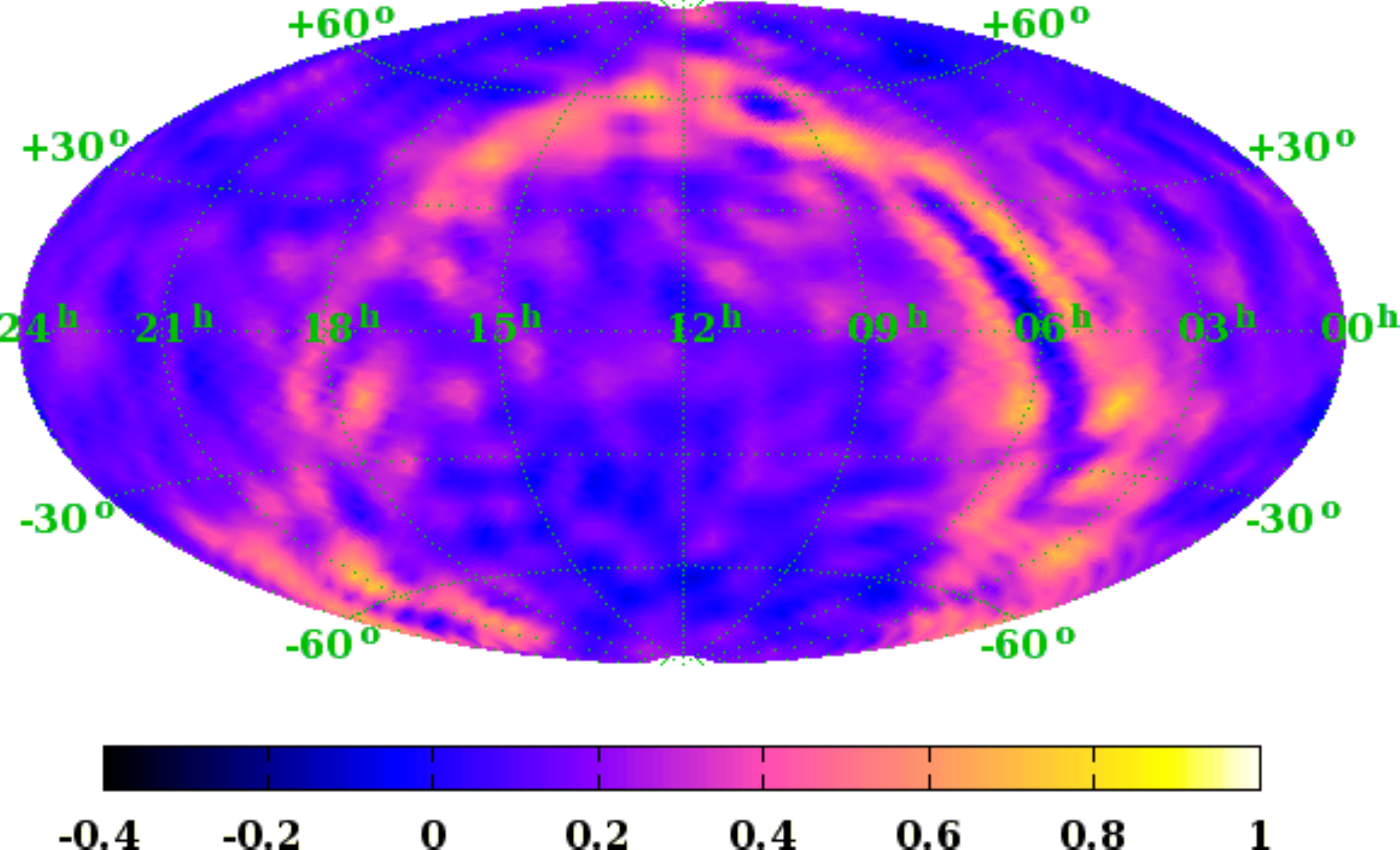}}\\
\subfigure[~Masked clean maps - obtained by putting a $4\sigma$ threshold on the clean maps\newline(left panel: NMSE = $0.36$; right panel: NMSE = $0.33$)]{\label{fig:DiffMap-d}%
\includegraphics[width=0.415\textwidth]{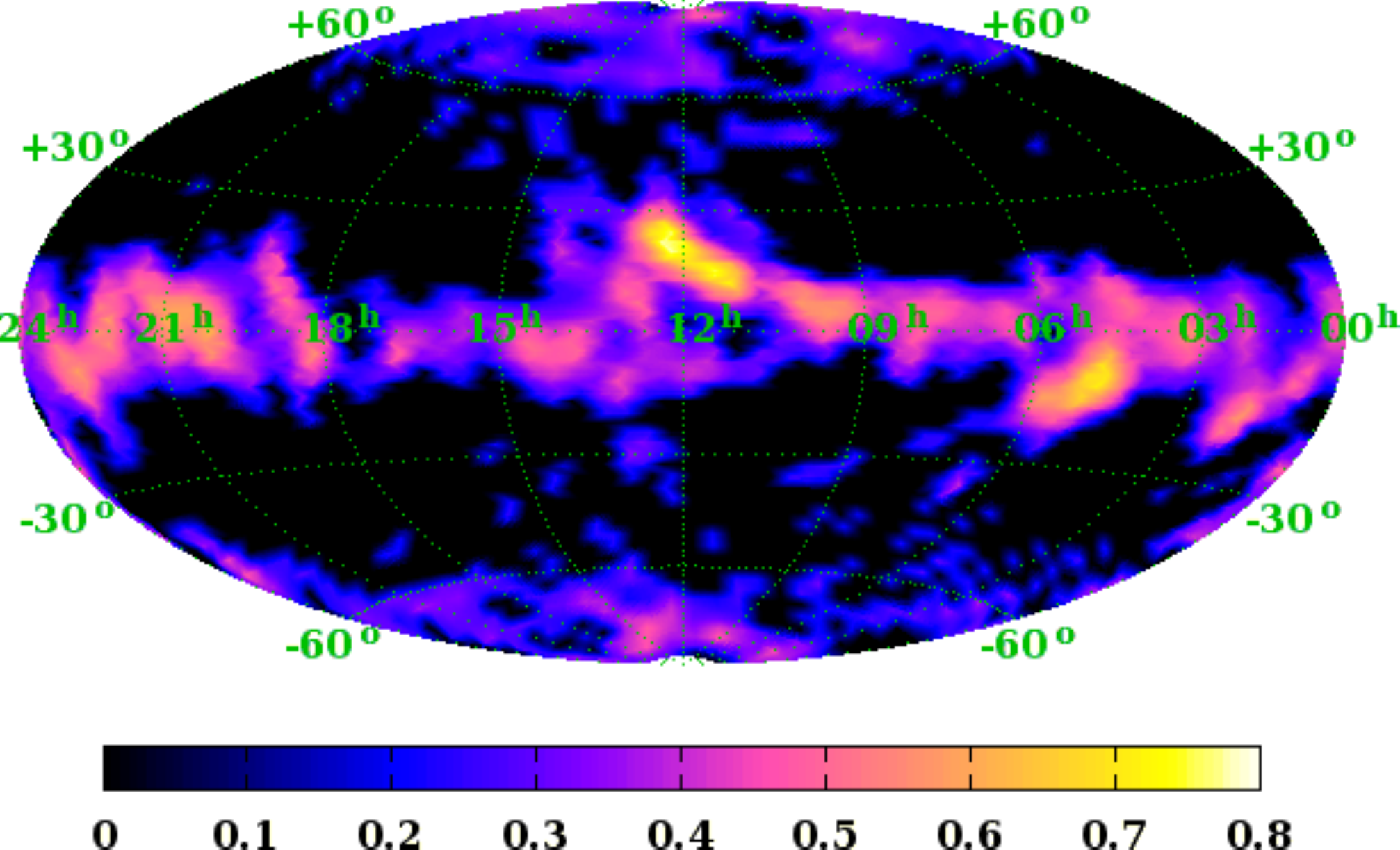}\qquad\qquad%
\includegraphics[width=0.415\textwidth]{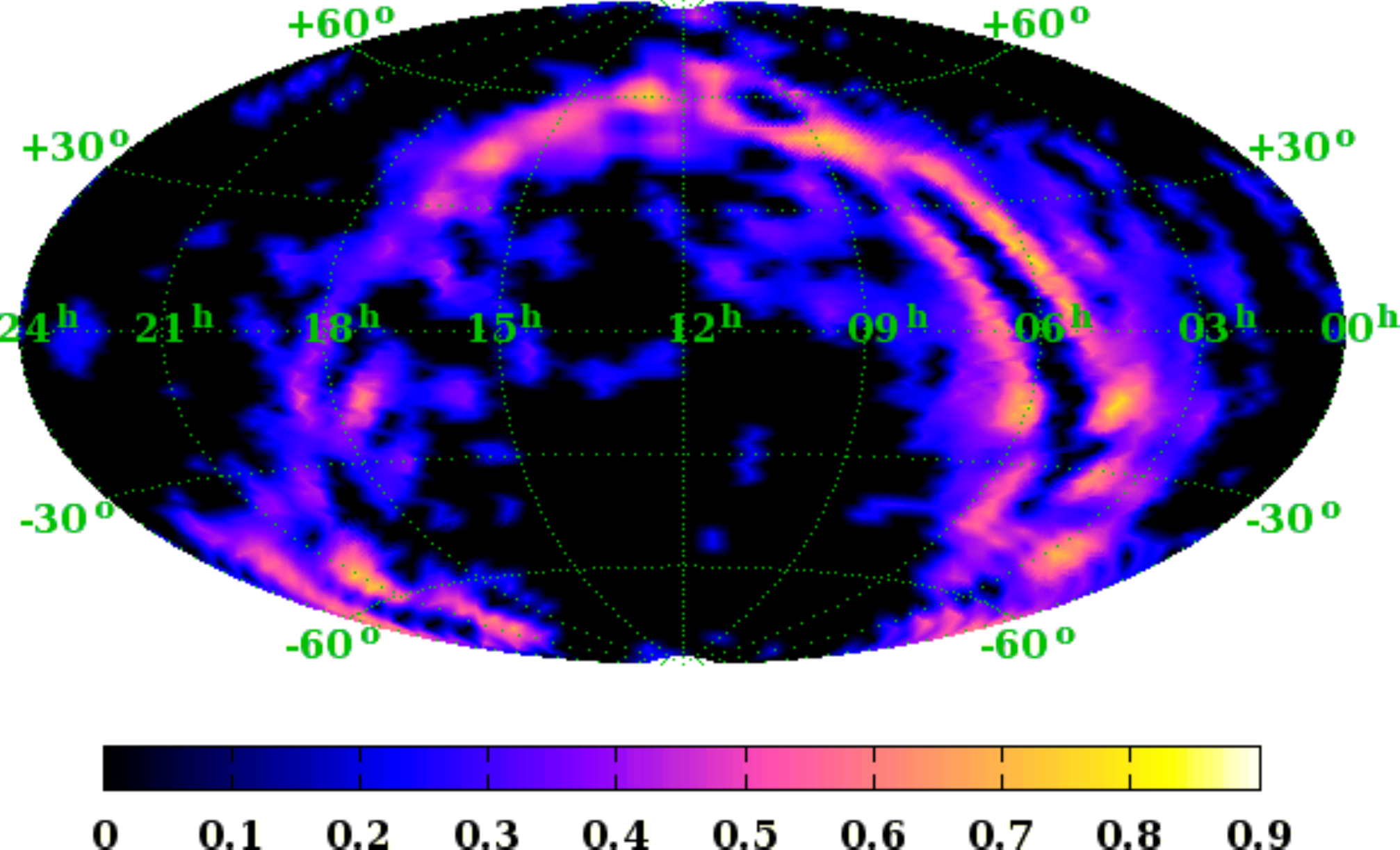}}
\caption{Illustration of deconvolution for broad sources: Maps similar to CMB temperature anisotropy sky with the galactic foreground were injected as toy maps. The left panels correspond to a map measured by the WMAP satellite~\cite{hin06}, as seen in the Galactic coordinates, as a toy model for an equatorial source and the right panels correspond to a map generated by the Planck Simulator~\cite{PlanckSim}, as a toy model for a multi-declination extended source.}
\label{fig:DiffMap}
\end{figure*}
%
Next, we inject two kinds of diffuse sources, viz., one that is nearly equatorial and another that is distributed across multiple declinations, as illustrated in Figure~\ref{fig:DiffMap}. We injected modified (using FTOOLS\footnote{\url{http://heasarc.gsfc.nasa.gov/ftools/}}) galactic foreground seen in CMB temperature anisotropy measurements as our test patterns for the diffuse SGWB sources. The left panels of Figure~\ref{fig:DiffMap} correspond to a modified form of the temperature anisotropy map measured by the WMAP satellite~\cite{hin06}.  We emphasise that the sky looks different in barycentric coordinates, similar to what is shown in the top-right panel of Figure~\ref{fig:DiffMap}. We omit the coordinate transformation step intentionally in order to get a diffuse equatorial source. The right panels of Figure~\ref{fig:DiffMap} correspond to a modified form of the temperature anisotropy map in the barycentric coordinates generated by the Planck Simulator~\cite{PlanckSim}. One of the main modifications applied to both of these maps was to mask the brightest part of the galaxy. This step reveals more structures in the maps, which is useful for testing a deconvolution algorithm. Figure~\ref{fig:DiffMap-a} shows the injected toy maps. Figure~\ref{fig:DiffMap-b} shows the dirty maps obtained by the radiometer analysis. One can see that the dirty maps have lost all the fine structures present in the injected maps. Furthermore, they show certain features that were not even present in the injected map. Also, the pixel values in the dirty maps are spread over a range consisting of large positive and almost equally negative values. Figure~\ref{fig:DiffMap-c} shows the clean maps recovered by $40$ CG iterations. Clearly, many of the features of the injected maps have been recovered in the clean maps, which is also evident from the lower values of NMSE, $0.33$ and $0.22$ respectively. Also, the positivity of the estimated map has been vastly improved - the pixel values of the clean maps lie mostly on the positive side, as one should expect. Finally, Figure~\ref{fig:DiffMap-d} shows the masked clean maps obtained by using a $4\sigma$ threshold. Though the masked maps give better visual impression, masking can actually discard several pixels which have weak sources, thereby increasing the NMSE. In Figure~\ref{fig:DiffMap-d}, for example, masking increases NMSE to $0.36$ and $0.33$ respectively, though the masked maps look more similar to the injected maps shown in Figure~\ref{fig:DiffMap-a}, than the unmasked clean maps in Figure~\ref{fig:DiffMap-c}.

\section{Conclusion}
\label{remarks}

The stochastic astrophysical GW background is likely to be dominated by sources in the nearby anisotropic universe, so the detection of localized sources is more favorable than the all-sky-averaged search. Making a skymap of the SGWB sky has been a long
standing ambition of stochastic GW research. Different analysis
methods have been proposed to create skymaps by measuring the first
few spherical harmonic multipoles of the sky. Here we have presented a  
direct approach of directed GW radiometer analysis. In this approach, the whole
sky is decomposed in a discrete set of pixels and the contribution
from each pixel is measured separately by correlating phase shifted
detector outputs to generate the whole skymap, which is a clear application of
earth rotation aperture synthesis. Specifically, for the AGWB detection statistic, we have defined a correlation statistic with a directed optimal filter that targets a fixed point in the sky by adjusting the time-delay across a baseline to track its rotation with the Earth. This statistic however provides us with a dirty map of the sky which we numerically deconvolve to obtain the true skymap. For this purpose, we have employed the conjugate gradient method. We numerically implement the deconvolution on simulated unpolarised GW sky maps obtained with the LIGO detector baseline. The success of this method is demonstrated by the recovery of simulated source distributions, namely, (i)~of a point source, (ii)~of a diffuse source in the equatorial plane, and (iii)~of a diffuse source off the equatorial plane.

This work needs to be implemented on other baselines of the upcoming/future network of detectors such as LIGO-VIRGO, LIGO-LCGT, LISA etc. The outline of the analysis has been presented here. However, further detailed analysis and the implementation is a future goal. Even for the single baseline, SPA analysis shows that, perhaps a more efficient method of deconvolution, yielding better accuracy and convergence, lower computational costs and convenience of application may be possible using more sophisticated analysis, for example, one involving basis functions. The Maximum Likelihood framework presented here is, in fact, independent of any particular choice of basis. So once a suitable basis is chosen, rest of the analysis can be applied without requiring any major change. It may also be possible to deconvolve only a patch of the sky using a similar method~\cite{burigana03}.

The work presented in this paper should also benefit two other searches. First, since
the long-duration integration of the data will essentially comprise a sum
over short stretches, a large signal in a short stretch will constitute
a candidate for a transient or burst (short-duration) event. Unlike 
the coincidence search being currently conducted for such events, our work
will combine coherently the outputs of several detectors and, thus,
improve their detectability. Second, the long-duration integration 
of the data should be able to find gravitational wave signals from
modelled sources, such as pulsars. Although our method is optimal
for searching unmodelled sources, it is not so for pulsars, the signals
from which can be matched filtered. The problem with the latter method is
that owing to the very large parameter space volume, an all-sky, all-frequency
search for pulsars with matched filtering is not computationally viable. 
Our proposed method is not handicapped by this problem since it does not use
the intrinsic source parameters for the search; rather it uses the data from one detector
to `filter' that from others in the network, after appropriately time-shifting
them and weighting them with the respective antenna patterns. Thus,
our method can be used as the first step in a two-step hierarchical search
for pulsars, where triggers from our method are followed up with 
matched filtering.

\section{Acknowledgments}

S. Mitra would like to acknowledge Council of Scientific and Industrial Research (India) and  Centre National d'Etudes Spatiales (France) for supporting his research and Caltech for supporting his visit to LIGO Laboratory, Caltech in 2006, where part of this work was done. He further thanks Stuart Anderson for helpful suggestions and Kent Blackburn and Patrick Sutton for providing useful help with the software computing facilities. Some of the results in this paper have been derived using the HEALPix~\cite{HEALPix} package, the Planck Simulator~\cite{PlanckSim} and FTOOLS~\cite{FTOOLS} and some of the plots were made using a colormap, specifically designed for compatibility with grayscale printing, included in the GNUPLOT package~\cite{gnuplot}. The LIGO Data Analysis System (LDAS) at Caltech and the High Performance Computing (HPC) facility at IUCAA were used for the numerical implementation. This work was supported in part by the Department of Science and Technology grant DST/INT/(US-RP077)/2001 and the National Science Foundation Grants INT-01-38459, PHY-0630121, PHY-0239735 and  the NSF LIGO Laboratory Cooperative Agreement PHY-0107417. This paper has LIGO Document Number LIGO-P070033-Z.



\end{document}